\pdfoutput=1  %% arxiv: uncomment !!!

\documentclass[preprints,article,accept,moreauthors,pdftex]{Definitions/mdpi}
\firstpage{1}      
\pubvolume{xx}
\issuenum{1}
\articlenumber{5}
\pubyear{2019}
\copyrightyear{2019}
\history{}  %% arXiv: use this, instead  %% !!!
\Title{Thermodynamical Extension of a Symplectic Numerical Scheme with Half
Space and Time Shift Demonstrated on Rheological Waves in Solids}

\Author{Tam\'as F\"ul\"op $^{1,2}$*\orcidA{},
R\'obert Kov\'acs $^{1,2,3}$\orcidB{},
M\'aty\'as Sz\"ucs $^{1,2}$\orcidC{} and Mohammad Fawaier $^{1}$}

\AuthorNames{Tam\'as F\"ul\"op, R\'obert Kov\'acs, M\'aty\'as Sz\"ucs
and Mohammad Fawaier}

\address{%
$^{1}$ \quad Department of Energy Engineering, Faculty of Mechanical
Engineering, BME, 1521 Budapest, Hungary; kovacsrobert@energia.bme.hu
(R.K.); szucsmatyas@energia.bme.hu (M.S.); fawaier.mohammad@gmail.com (M.F.)
 \\
$^{2}$ \quad Montavid Thermodynamic Research Group, 1112 Budapest, Hungary;
 \\
$^{3}$ \quad Department of Theoretical Physics, Wigner Research Centre for
Physics, Institute for Particle and Nuclear Physics, 1525 Budapest, Hungary}

\corres{Correspondence: fulop@energia.bme.hu}

\abstract{%
On the example of the \PTZ\ rheological model for
solids, which exhibits both dissipation and wave propagation -- with
nonlinear dispersion relation --, we introduce and investigate a finite
difference numerical scheme.
{Our goal is to demonstrate its properties and to ease the
computations in later applications for continuum thermodynamical
problems.}
The key element is the positioning of the discretized quantities with shifts
by half space and time steps with respect to each other. The arrangement
is chosen according to the spacetime properties of the quantities and of
the equations governing them. Numerical stability, dissipative error and
dispersive error are
 analysed 
in detail. With the best settings found,
the scheme is capable of making
precise and fast predictions.
{Finally, the proposed scheme is compared to a commercial
finite element software, COMSOL, which demonstrates essential differences
even on the simplest -- elastic -- level of modelling.}
 \OMIT{%
A single paragraph of about 200 words maximum. For research
articles, abstracts should give a pertinent overview of the work. We
strongly encourage authors to use the following style of structured
abstracts, but without headings: (1) Background: Place the question
addressed in a broad context and highlight the purpose of the study; (2)
Methods: Describe briefly the main methods or treatments applied; (3)
Results: Summarize the article's main findings; and (4) Conclusion:
Indicate the main conclusions or interpretations. The abstract should be
an objective representation of the article, it must not contain results
which are not presented and substantiated in the main text and should
not exaggerate the main conclusions.
 }%(V)OMIT
}

 %% 3--10 needed
\newcommand*\nc{\newcommand*}  \nc\longnc{\newcommand}
\longnc\VOMIT[1]{#1}  \longnc\OMIT[1]{}  %% swithcing on/off manuscript parts

 \VOMIT{%  %% !!!
\nc\ig[2]{\centerline{\includegraphics[#1]{Figures/#2}}}
\nc\igg[3]{\centerline{\hfill\includegraphics[#1]{Figures/#2}\hfill
  \includegraphics[#1]{Figures/#3}\hfill}}
\nc\iggg[4]{\centerline{\includegraphics[#1]{Figures/#2}\hfill
  \includegraphics[#1]{Figures/#3}\hfill\includegraphics[#1]{Figures/#4}}}
\nc\igggg[5]{\centerline{\includegraphics[#1]{Figures/#2}\hfill
  \includegraphics[#1]{Figures/#3}\hfill
  \includegraphics[#1]{Figures/#4}\hfill\includegraphics[#1]{Figures/#5}}}
}%(V)OMIT

\nc\subheader{\m { \qldskw{} , \qlsskw{} } in \m { \qMpos } &
  \m { \qldskw{} , \qlsskw{} } in \m { \qMmin } }

\nc\lat\textit  %% latin: i.e. e.g. in situ 
\nc\ie{\lat{i.e.,\ }}  \nc\etal{\lat{et al.}}    \nc\etc{\lat{etc.\ }}
\nc\eg{\lat{e.g.,\ }}  \nc\ergo{\lat{ergo}}
\nc\cf{cf.\ }
\nc\lhs{lhs}      \nc\rhs{rhs}
\nc\PTZ{Poynting--Thomson--Zener}

\nc\mathBf{\boldsymbol}

\nc\m[1]{$ #1 $}                     %% space around an in-text formula
\nc\mm[1]{\m{ \, #1 \, }}            %% Tip: always use \m{...}
\nc\mmm[1]{\m{ \,\, #1 \,\, }}       %% instead of $...$, thus later
\nc\mmmm[1]{\m{ \,\,\, #1 \,\,\, }}  %% you can add space easily.
\nc\M[3]{\scase=0$                 %% finer and unequal spaces
	\mskip#1mu#3\mskip#2mu$}       %% around an in-text formula
\nc\re[1]{(\ref{#1})}

\nc\Mbox[1]{\makebox[19.6em][l]{$\displaystyle #1$}}
\nc\Mboxx[1]{\makebox[14.4em][l]{$\displaystyle #1$}}
\nc\Mboxxx[1]{\makebox[17em][l]{$\displaystyle #1$}}
\nc\Mboxxxx[1]{\makebox[15em][l]{$\displaystyle #1$}}

\nc\0[2]{\ifcase#1{#2}\or\lt(#2\rt)\or\lt[{#2}\rt]\or\lt\{{#2}\rt\}\or
  \mathord<{#2}\mathord>\or\lt\langle{#2}\rt\rangle\or\lt\lvert{#2}\rt
  \rvert\or\lt\lVert{#2}\rt\rVert\fi}
\nc\1[2]{\ifcase#1{#2}\or(#2)\or[#2]\or\{#2\}\or\mathord<{#2}\mathord
  >\or\langle{#2}\rangle\or\lvert{#2}\rvert\or\lVert{#2}\rVert\fi}
\nc\2[2]{\ifcase#1{#2}\or\big(#2\big)\or\big[#2\big]\or\big
  \{#2\big\}\or\big<#2\big>\or\big\langle#2\big\rangle\or\big
  \lvert#2\big\rvert\or\big\lVert#2\big\rVert\fi}
\nc\3[2]{\ifcase#1{#2}\or\Big(#2\Big)\or\Big[#2\Big]\or\Big\{#2\Big
  \}\or\Big<#2\Big>\or\Big\langle#2\Big\rangle\or\Big\lvert#2\Big
  \rvert\or\Big\lVert#2\Big\rVert\fi}
\nc\4[2]{\ifcase#1{#2}\or\bigg(#2\bigg)\or\bigg[#2\bigg]\or\bigg
  \{#2\bigg\}\or\bigg<#2\bigg>\or\bigg\langle#2\bigg\rangle\or\bigg
  \lvert#2\bigg\rvert\or\bigg\lVert#2\bigg\rVert\fi}
\nc\5[2]{\ifcase#1{#2}\or\Bigg(#2\Bigg)\or\Bigg[#2\Bigg]\or\Bigg
  \{#2\Bigg\}\or\Bigg<#2\Bigg>\or\Bigg\langle#2\Bigg\rangle\or\Bigg 
  \lvert#2\Bigg\rvert\or\Bigg\lVert#2\Bigg\rVert\fi}
\nc\9[2]{\ifcase#1{#2}\or\left(#2\right)\or\left[#2\right]\or\left
  \{#2\right\}\or\left\langle{#2}\right\rangle\or\left\langle{#2}\right
  \rangle\or\left\lvert{#2}\right\rvert\or\left\lVert{#2}\right\rVert\fi}
\nc\lt{\mathopen{}\mathclose\bgroup\left} \nc\rt{\aftergroup\egroup\right}

\nc\bi\relax                     %% spacing finer than \! \, \: \;
\nc\bit{    \mskip1mu}  \nc\biT{    \mskip-1mu} %% Note:
\nc\bitt{   \mskip2mu}  \nc\biTT{   \mskip-2mu} %% \! = -3mu,
\nc\bittt{  \mskip3mu}  \nc\biTTT{  \mskip-3mu} %% \, = 3mu,
\nc\bitttt{ \mskip4mu}  \nc\biTTTT{ \mskip-4mu} %% \: = 4mu,
\nc\bittttt{\mskip5mu}  \nc\biTTTTT{\mskip-5mu} %% \; = 5mu.

\nc\f\frac   \nc\ff[2]{{\textstyle \f{#1}{#2}}}  \nc\F[2]{#1/#2}
\nc\fF{\ff}  \nc\fFF{\f}
\nc\restr[2]{{\lt.#1\rt|}_{#2}} %% restriction; value at #2
\nc\tr{\operatorname{tr}}

\nc\e{\mathrm{e}} %% \e = 2.718281828459...
\nc\ii{\mathrm{i}} %% \ii^2 = -1
\nc\dd{\mathrm{d}} \nc\ddd{\bit\d} %% differential d
\nc\pd\partial
\nc\var\delta

\nc\doubledot{:}
\nc\singledot{\cdot}
\nc\dyad{\otimes}
\nc\Trans{\text{T}}
\nc\tensor\mathbf  \nc\Tensor\mathBf  \nc\tensoR\mathsf
\nc\sym[1]{{#1}^\text{S}}              \nc\skw[1]{{#1}^\text{A}}
\nc\dev[1]{{#1}^\text{d}}              \nc\sph[1]{{#1}^\text{s}}

\nc\devv[1]{{#1}\bit^\text{d}}         \nc\sphh[1]{{#1}\bit^\text{s}}
\nc\tot[1]{{#1}_\text{total}}           \nc\kin[1]{{#1}_\text{kinetic}}
\nc\ther[1]{{#1}_\text{thermal}}        \nc\ela[1]{{#1}_\text{elastic}}
\nc\rheol[1]{{#1}_\text{rheological}}
\nc\dyn[1]{{#1}_\text{dyn}}
\nc\aux[1]{{#1}_\text{aux}}
\nc\nd[1]{\tilde{#1}}  %% dimensionless version of quantities
\nc\qinner{\Delta}
\nc\Quad{\:\:}
\nc\qdot[1]{\f {\pd #1}{\pd \qt}}  \nc\Pri[1]{\f {\pd #1}{\pd \qx}}
\nc\RE{\operatorname{Re}}          \nc\IM{\operatorname{Im}}
\nc\Ordo{\mathcal{O}}
\nc\qqh[1]{#1}  \nc\qqp[1]{\hat{#1}}
\nc\Qa{a}              \nc\qai[1]{\Qa_{#1}}
\nc\qb{b}              \nc\qa[1]{\qb_{#1}}
\nc\qc{c}              \nc\qcc{\qqp{\qc}}    \nc\qcp{c_{\qsig}}
\nc\qcC{\f {\qEY}{\qrho}}  \nc\qcCF{\fF {\qEY}{\qrho}}
\nc\qe{e}
\nc\qj{j}  \nc\qjm{\qj - \F12}    \nc\qjp{\qj + \F12}
\nc\qk{k}
\nc\qldsym[1]{l^\text{d}_\text{S}}     \nc\qlssym[1]{l^\text{s}_\text{S}}
\nc\qldskw[1]{l^\text{d}_\text{A}}     \nc\qlsskw[1]{l^\text{s}_\text{A}}
\nc\qlsym{l_\text{S}}                  \nc\qlskw{l_\text{A}}
\nc\qm{m}
\nc\qn{n}  \nc\qnm{\qn - \F12}    \nc\qnp{\qn + \F12}
\nc\qs{s}
\nc\qt{t}  \nc\qDt{\Delta \qt}
\nc\qv{v}
\nc\qx{x}  \nc\qDx{\Delta \qx}
\nc\qy{\tensoR y}

\nc\qAv{A_{\qv}}       \nc\qAvv[1]{\qAv\01{#1}}
\nc\qAeps{A_{\qeps}}   \nc\qAepss[1]{\qAeps\01{#1}}
\nc\qAsig{A_{\qsig}}   \nc\qAsigg[1]{\qAsig\01{#1}}
\nc\qAAsig[1]{A_{\qsig,#1}}
\nc\qC{C}             \nc\qCC{\qqp{\qC}}
\nc\qEY{E}              \nc\qEE{\hat{\qEY}}     \nc\qEEE{\hat{\hat{\qEY}}}
\nc\qEinf{\qEY_\infty}
\nc\qH{H}
\nc\qII{\hat{I}}
\nc\qJ{J}
\nc\qN{N}
\nc\qP{P}              \nc\qPP{\qqp{\qP}}
\nc\qQ{Q}              \nc\qQQ{\qqp{\qQ}}
\nc\qS{S}
\nc\qSS{\sin{\f{\qk \qDx}{2}}}
\nc\qT{T}              \nc\qqT{\tensoR T}     \nc\qqTT{\qqp{\qqT}}
\nc\qX{X}

\nc\qalp{\alpha}
\nc\qbet{\beta}
\nc\qeps{\varepsilon}  %\nc\qdeps{\dot{\qqeps}}
\nc\qrho{\varrho}
\nc\qsig{\sigma}       \nc\qsigb{\qsig_\text{b}}
\nc\qtau{\tau}         \nc\qtauu{\hat{\qtau}}
\nc\qtaub{\tau_\text{b}}  \nc\qndtaub{{\nd{\tau}}_\text{b}}
\nc\qxi{\xi}  \nc\qxii{\qxi_+}  \nc\qxiii{\qxi_-}  \nc\qxiiii{\qxi_\pm}
\nc\qeta{\eta}
\nc\qome{\omega}

\nc\qFIG{22.5ex}
\nc\qFig{16.ex}  %% \xi
\nc\qCOMSOL{.4\textwidth}

\nc\Red{}
\nc\RRed{}

\begin{document}  %%%%%%%%%%%%%%%%%%%%%%%%%%%%%%%%%%%%%%%%%%

%\newpage  %% !!!

\section{Introduction}

Numerical solution methods for dissipative problems are
important and are a nontrivial topic. Already for reversible systems,
the difference between a symplectic and nonsymplectic finite difference
method is striking: the former can offer reliable prediction that stays
near the exact solution even at extremely large time scales while the
latter may provide a solution that drifts away from the exact solution
%vigorously.
 steadily.
For dissipative systems, the situation is harder. Methods that were born
with reversibility in mind may apparently fail for a nonreversible
problem. For example, a finite element software is able to provide, at
the expense of large run time, quantitatively and even qualitatively
wrong outcome while a simple finite difference scheme solves the same
problem fast and
% \RRed{accurately}
 precisely
\cite{rieth-kovacs-fulop:2018}.

Partly inspired by, and partly based on, the intensive development on
symplectic schemes for reversible problems, remarkable
research is done in recent years to develop geometric methods
for dissipative systems, more on ones with finite
degrees of freedom \1 1 {including
\cite{zinner-ottinger:2019,shang-ottinger:2018,portillo-etal:2019%
 ,vermeeren-etal:2019,gayos18,couga20}}
and less for continua \1 1 {see, \eg
\cite{romero1:2010,romero2:2010,BerVan17b,czech}}.
% \Red{We also note that symplectic methods are mostly geometrically
%motivated: the phase-volume (or phase-area) preservation along the time
%evolution must be respected by the numerical method itself.}

Thermodynamics also modifies the way of thinking concerning numerical
%\Red{modeling}.
%% British English also OK: https://www.mdpi.com/authors/english-editing
 modelling.
Even if quantities known from mechanics form a closed system
of equations to solve numerically, monitoring temperature \1 1 {or other
thermodynamical quantities} for a nonreversible system can give insight
on the processes and phenomena, for example, pointing out the presence
of viscoelasticity/rheology, and displaying when plastic changes start
\cite{asszonyi-csatar-fulop:2016}. In addition,
% \Red{the}  %% !!! overruling Grammarly
 temperature can also
react, in the form of thermal expansion and heat conduction, even in
situations where one is not prepared for this `surprise'
\cite{emergence}.

Furthermore, in a sense, thermodynamics is a stability theory.
Therefore, how thermodynamics ensures asymptotic stability for systems
may give new ideas on how stability and suppression of errors can be
achieved for numerical methods. A conceptually closer relationship is
desirable between these two areas.

Along these lines,
% here,
we present a study where a new numerical scheme is suggested and applied
for a continuum thermodynamical model. The scheme proves to be an
extension of a symplectic method. In parallel, our finite difference
scheme introduces a \Red{staggered} arrangement of quantities by half space and
time steps with respect to each other, according to the spacetime nature
of the involved quantities and the nature of equations governing them.
The shifts can be introduced by inspecting the equations. It turns out
that balances, kinematic equations, and Onsagerian equations all have
their own distinguished discretized realization. The shifts also make
the scheme one order higher \Red{accuracy} as the original symplectic scheme.

The continuum system
that
we take as the subject of our investigation is important on its own --
it is the \PTZ\ \Red{(PTZ)} rheological model for solids. This
model exhibits
{both} dissipation and wave propagation \1 1 {actually:
dispersive wave propagation}, and
 this
is thus ideal for testing various
aspects and difficulties. Meanwhile, its predictions are relevant for
many solids, typically ones with complicated micro- or mesoscopic
structure like rocks \cite{lin-etal:2010,matsuki:1993,matsuki:2008},
plastics \cite{asszonyi-csatar-fulop:2016}, asphalt \etc This
non-Newtonian rheological model can explain why slow and fast
measurements and processes give different results.

Solutions in the force equilibrial and space independent limit have
proved successful in explaining experimental results
\cite{asszonyi-csatar-fulop:2016}. Space dependent -- but still force
equilibrial -- analytical solutions can model opening of a tunnel,
gradual loading of thick-walled tubes and spherical tanks, and other
problems \cite{fulop-szucs:2018}. The next level is to leave the force
equilibrial approximation, partly in order to cover and extend the force
equilibrial results but also to be utilized for evaluating measurements
that include wave propagation as well. The present work is, in this
sense, the next step in this direction.

\section{Properties of the continuum model}

The system that we consider is a homogeneous solid with
Poynting--Thomson--Zener
% \1 1 {PTZ}
rheology, in small-strain approximation\footnote{Hence, there is no need
to distinguish between Lagrangian and Eulerian variables, and between
material manifold vectors/covectors/tensors/\ldots and spatial spacetime
ones.}, in one space dimension \1 1 {1D}. Notably, the numerical scheme
we introduce in the following section can be generalized to 2D and 3D
with no difficulty\footnote{The results of our ongoing research on 2D
and 3D are to be communicated later.} -- the present 1D treatment is to
keep the technical details at a minimum so we can focus on the key
ideas.

The set of equations we discuss is, accordingly,
 \begingroup
\addtolength{\jot}{1ex}  %% !!!
 \begin{align}  \label{A}
\qrho \qdot{\qv} & = \Pri\qsig ,
 \\  \label{B}
\qdot{\qeps} & = \Pri\qv,
 \\  \label{C}
\qsig + \qtau \qdot{\qsig} &= \qEY \qeps + \qEE \qdot{\qeps} ,
 \end{align}
 \endgroup
where \m { \qrho } is mass density \1 1 {constant in the small-strain
approximation}, \re{A} tells how the spatial derivative of
stress \m { \qsig } determines the time derivative of the velocity field
\m { \qv } \1 1 {volumetric force density being omitted for simplicity},
\re{B} is the kinematic relationship between the strain field \m
{ \qeps } \footnote{In the present context, \m { \qeps } can be used as
the thermodynamical state variable for elasticity, but not in general,
see \cite{fulop-van:2012,fulop:2015}.} and \m { \qv }, and the rheological
relationship \re{C} contains, in addition to Young's modulus \m {
\qEY }, two
%non-negative -- in what follows, considered positive if not mentioned
%otherwise --
 positive
coefficients \m { \qEE , \qtau}.

The \Red{PTZ} model is a
% special
subfamily within the Kluitenberg--Verh\'as model family, which family
can be obtained via a nonequilibrium thermodynamical internal variable
approach \cite{asszonyi-fulop-van:2015}.
%In case of the \PTZ\ model, the result reached by eliminating the
%internal variable is particularly simple both in specific internal
%energy \m { \qe } and in specific entropy \m { \qs }:
The \Red{PTZ} model looks particularly simple, after eliminating the
internal variable, both in specific energy \m { \tot\qe } and in
specific entropy \m { \qs }:
 \begin{align}  \label{energy}
\tot\qe & = \kin\qe + \ther\qe + \ela\qe + \rheol\qe \equiv
\f {1}{2} \qv^2 + \qcp \qT + \f {\qEY}{2 \qrho} \qeps^2 +
\f {\qtau}{2 \qrho \qII} \9 1 { \qsig - \qEY \qeps }^2 ,
 \\  \label{entropy}
\qs & = \qcp \ln \f {\qT}{\aux\qT}
 \end{align}
\1 1 {%
%via
along the lines of
\cite{asszonyi-fulop-van:2015}, Appendix B%
%, with details given in \cite{fulop:2019}%
}, where thermal
expansion and heat conduction are neglected and the `isobaric' specific
heat \m { \qcp } is assumed constant for simplicity, \m { \qT } is
absolute temperature, the auxiliary constant \m { \aux\qT } is present
on dimensional grounds, and
  the `index of damping' \m { \qII } \cite{asszonyi-fulop-van:2015} is
 \begin{align}  \label{ineq}
\qII = \qEE - \qtau \qEY > 0 ,
 \end{align}
the inequality being a consequence of the second law of thermodynamics.
Moreover, in this simple setting, entropy production rate density,
 \begin{align}  \label{entprod}
\f {1}{\qT} \f {\9 1 { \qsig - \qEY \qeps }^2}{\qII} ,
 \end{align}
increases temperature directly:\footnote{%
Eq.~\re{Tdot} can be understood directly by taking
%This is directly visible from
\m { \qrho \qT } times the time derivative of \re{entropy},
% also bearing in mind
 together with the balance of entropy and the fact
that, with neglected heat conduction, entropy current density has also
been set to zero.}
 \begin{align}  \label{Tdot}
\qrho \qcp \qdot\qT = \f {\9 1 { \qsig - \qEY \qeps }^2}{\qII} .
 \end{align}

Remarkably,\footnote{And thanks to our simplifications.} the closed
system of equations \re{A}--\re{B}, \re{C} to be solved is linear.
Having the solution for \m { \qv }, \m { \qeps }, and \m { \qsig }, the
further quantities \1 1 {\m { \qT }, \m { \qs }, and the various energy
terms} can be obtained.

Our system admits two distinguished time scales, \m { \qtau } and
 \begin{align}  \label{tauhat}
\qtauu = \f {\qEE}{\qEY} > \qtau ,
 \end{align}
the inequality following from \re{ineq}. For phenomena much slower than
these time scales, the rule-of-thumb approximation of keeping only the
lowest time derivative for any quantity present in \re{C} gives the
Hooke model
 \begin{align}  \label{hooke}
\qsig = \qEY \qeps ,
 \end{align}
formally the \m { \qtau \to 0, } \m { \qEE \to 0 } \1 1 {%
%or
%in other words,
\m {\qtauu \to 0}} limit of \re{C}. The system of equations
\re{A}--\re{B}, \re{hooke} leads to a wave equation for \m { \qv }, \m {
\qsig }, \m { \qeps } each, with wave speed
 \begin{align}  \label{c}
\qc = \sqrt{ \f {\qEY}{\qrho} } .
 \end{align}
On the other side, for processes much faster then the two time scales,
keeping the highest time derivatives leads to
 \begin{align}  \label{fast}   
\qtau \qdot{\qsig} &= \qEE \qdot{\qeps} ,
 &
\qdot{\qsig} &= \f{\qEE}{\qtau} \qdot{\qeps} ,
 & & \Longrightarrow \; \int_{\qt_1}^{\qt_2} \dd \qt \;\, \text{gives} &
\Delta_{\qt_1 \to \qt_2}^{} \qsig & = \f{\qEE}{\qtau} \Delta_{\qt_1 \to
\qt_2}^{} \qeps ,
 \end{align}
that is, for stress and strain changes \1 1 {\eg for deviations from
initial values}, the system effectively behaves like a Hooke one, with
`dynamic' Young's modulus
 \begin{align}  \label{Edyn}
\qEinf = \f{\qEE}{\qtau} ,  \qquad \quad  \qEinf > \qEY .
 \end{align}
The corresponding effective wave equation possesses the wave speed
 \begin{align}  \label{cc}
\qcc = \sqrt{ \f {\qEinf}{\qrho} } = \sqrt{ \f {\qEE}{\qtau\qrho} },
 \qquad \quad
\qcc > \qc  .
 \end{align}

For a more rigorous and closer investigation of these aspects, the
dispersion relation can be derived. Namely, on the line \m { - \infty <
\qx < \infty }, any \1 1 {not too pathological} field can be given as a
continuous linear combination of \m { \e^{\ii \qk \qx} } space
dependences, where the `wave number' \m { \qk } is any real parameter.
%% \m{\ii=\sqrt{-1}} is the imaginary unit
If such a \1 1 {Fourier} decomposition is done at, say, \m { \qt = 0 },
then the subsequent time dependence of one such mode may be particularly
simple:
% written together for our three fields
 \begin{align}  \label{mode}
\begin{pmatrix} \qv \\ \qeps \\ \qsig \end{pmatrix} \1 1 {
%\qt = 
0, \qx} =
\begin{pmatrix} \ii \qAvv{\qk} \\ \qAepss{\qk} \\ \qAsigg{\qk} \end{pmatrix}
\e^{\ii \qk \qx} ,  \qquad \qquad
\begin{pmatrix} \qv \\ \qeps \\ \qsig  \end{pmatrix} \1 1 {\qt, \qx} =
\begin{pmatrix} \ii \qAvv{\qk} \\ \qAepss{\qk} \\ \qAsigg{\qk} \end{pmatrix}
\e^{-\ii \qome \qt} \e^{\ii \qk \qx}
 \end{align}
with some appropriate \m { \qome } -- complex, in general --; the factor \m { \ii } in the first
component is introduced in order to be in tune with later convenience.
A space and time dependence
 \begin{align}  \label{traveldec}
\e^{-\ii \qome \qt} \e^{\ii \qk \qx} = \e^{\IM \qome \qt}
\e^{-\ii \RE \qome \qt} \e^{\ii \qk \qx} = \e^{- \1 1 {-\IM \qome} \qt}
\e^{ \ii \qk \9 1 { \qx - \f {\RE \qome}{\qk} \qt} }
 \end{align}
expresses travelling with constant velocity \m { \f {\RE \qome}{\qk} }
and exponential decrease \1 1 {for dissipative systems like ours, \m {
\IM \qome < 0 }}. In general, it depends on the number of fields and on
the order of time derivatives how many \m { \qome }'s are possible. In
our case, the relationship between compatible \m { \qome } and \m { \qk
} -- the dispersion relation -- is straightforward to derive:
 \begin{align}  \label{disp}
\qome^2 \f {1 - \ii \qtau \qome}{1 - \ii \qtauu \qome} = \qc^2 \qk^2 .
 \end{align}
In the limit \m { \0 6 {\qome} \to 0 } \1 1 {limit of slow processes},
we find
 \begin{align}  \label{disph}
\qome^2 = \qc^2 \qk^2 ,  \qquad\quad  \qome = \pm \qc \qk ,  \qquad\quad
\f {\RE \qome}{\qk} = \f {\qome}{\qk} = \pm \qc ,
 \end{align}
while in the opposite limit \m { \0 6 {\qome} \to \infty } \1 1 {limit
of fast processes}, the result is
 \begin{align}  \label{dispinf}
\qome^2 \f {- \ii \qtau \qome}{- \ii \qtauu \qome} =
\qome^2 \f {\qtau}{\qtauu} = \qc^2 \qk^2 ,  \qquad\quad
\qome = \pm \qcc \qk ,  \qquad\quad
\f {\RE \qome}{\qk} = \f {\qome}{\qk} = \pm \qcc .
 \end{align}
Both results confirm the findings above \1 2 {\re{c} and \re{cc},
respectively}.

This is a point where we can see the importance of the \Red{PTZ} model.
Namely, when measuring Young's modulus \1 1 {or, in 3D, the two
elasticity coefficients} of a solid, the speed of uniaxial loading, or
the frequency of sound in an acoustic measurement, may influence the
outcome and an adequate/sufficient interpretation may come in terms of a
\Red{PTZ} model. Indeed, in rock mechanics, dynamic elastic moduli are long
known to be larger than their static counterparts \1 1 {a new and
comprehensive study on this, \cite{mortaza}, is in preparation}, in
accord with the thermodynamics-originated inequality in \re{cc} \1 1 {or
its 3D version%, \cite{jegyzet}
}.

\section{The numerical scheme}

The classic attitude to finite difference schemes is that all 
quantities are registered at the same discrete positions and at the same
discrete instants. An argument against this practice is that, when
dividing a sample into finite pieces, some physical quantities have a
meaning related to the bulk, the centre of a piece, while others have a
physical role related to the boundaries of a unit. For example, in
Fourier heat conduction, heat flux is proportional to the gradient of
temperature. A natural discretization of this, in one space dimension,
is that temperature values sit at the centres and heat flux values at
the boundaries -- in other words, at a half space step distance from the
centres \cite{rieth-kovacs-fulop:2018}. Also in heat conduction, change
rate of specific internal energy is determined by the divergence of the
heat flux. The natural one space dimensional discretization is then
that, since heat flux values sit at the boundaries, specific internal
energy values \m { \ther{\qe} } are placed at the centres \1 1 {at the
same places where temperature values \m { \qT } are put, which is in
tune with that the two are related to one another via \m { \ther{\qe} =
\qcp \qT }} \cite{rieth-kovacs-fulop:2018}. More generally in
% \Red{the}  %% !!! overruling Grammarly
continuum
theories, specific extensive and density quantities would naturally live
at a centre, while currents/fluxes are boundary related by their
physical nature/role.

Here, we generalize this approach. Namely, when one has a full -- at the
general level, 4D -- spacetime perspective\footnote{Traditional physical
quantities are usually time- and spacelike components of four-vectors,
four-covectors, four-cotensors \etc, which are governed by 4D equations
with four-divergences, four-gradients \etc} then it turns out that
quantities may ``wish'' to be \Red{staggered} with respect to each other by a
half in time as well. Taking again the example of the balance of
internal energy in heat conduction: the finite-difference discretization
of the change rate of specific internal energy \m { \ther{\qe} }
contains the
%difference
change \m { \Delta \ther{\qe} } corresponding to a finite time
difference \m { \qDt }.
%, divided by this time step \m { \qDt }.
This change is caused by the flux of heat leaving the spatial unit
during this time interval \m { \qDt }, the time average of the flux
naturally realized at half-time \m { \F {\qDt}{2} }. Accordingly, heat
flux values would be realized as half-shifted in time with respect to
specific internal energy.

More generally, if an equation relates the change rate of a quantity to
another quantity then these two quantities would be realized as
half-shifted in time with respect to one another.

To sum up, the space and time derivatives give us hints to arrange the
quantities with space and time half-shifts, respectively.
%The structure of the equations helps us to reveal what
%intends to be shifted with respect to what.

This approach is what we realize for the present system. Discrete space
and time values are chosen as
 \begin{align}  \label{disc}
\qx_{\qn} = \qn \qDx ,  \quad  \qn = 0, 1, \ldots, \qN , 
 \qquad
\qt^{\qj} = \qj \qDt ,  \quad  \qj = 0, 1, \ldots, \qJ ,
 \end{align}
and discrete values of stress are prescribed to these spatial and
temporal coordinates:
 \begin{align}  \label{sighol}
\qsig^{\qj}_{\qn} \quad \text{at} \quad \qt^{\qj}, \; \qx_{\qn} .
 \end{align}
Then, investigating \re{A}, we decide to put velocity values
half-shifted with respect to stress values both in space and
time:%
%% v^j at j-1/2, v_n at n+1/2
%\footnote{One could introduce a notation like \m {
%\qv^{\qj-1/2}_{\qn+1/2} } to emphasize this. Here, we stay with integer
%indices to help direct software implementation.}
 \begin{align}  \label{vhol}
\qv^{\qjm}_{\qnp} \quad \text{at} \quad
%\qt^{\qjm} \equiv
\qt^{\qj} - \ff {\qDt}{2}, \quad
%\qx_{\qnp} \equiv
\qx_{\qn} + \ff {\qDx}{2} ,
 \end{align}
and discretize \re{A} as
 \begin{align}  \label{Ad}
%\qrho \f { \qv^{\qj+1}_{\qn} - \qv^{\qj}_{\qn} }{\qDt}
\qrho \f { \qv^{\qjp}_{\qnp} - \qv^{\qjm}_{\qnp} }{\qDt}
& = \f { \qsig^{\qj}_{\qn+1} - \qsig^{\qj}_{\qn} }{\qDx} .
 \end{align}
Next, studying \re{B} suggests analogously to have strain values
half-shifted with respect to velocity values both in time and space.
Therefore, strain is to reside at the same spacetime location as stress:
 \begin{align}  \label{epshol}
\qeps^{\qj}_{\qn} \quad \text{at} \quad \qt^{\qj}, \; \qx_{\qn} ,
 \end{align}
and \re{B} is discretized as
 \begin{align}  \label{Bd}
\f { \qeps^{\qj+1}_{\qn} - \qeps^{\qj}_{\qn} }{\qDt}
%& = \f { \qv^{\qj+1}_{\qn} - \qv^{\qj+1}_{\qn-1} }{\qDx} .
& = \f { \qv^{\qjp}_{\qnp} - \qv^{\qjp}_{\qnm} }{\qDx} .
 \end{align}
Finally, for the Hooke model, \re{hooke} is discretized plainly as
 \begin{align}  \label{HCd}
\qsig^{\qj}_{\qn} = \qEY \qeps^{\qj}_{\qn}
 \end{align}
as stress and strain are
% attributed to
% allocated at
 assigned to
the same locations. Actually, in the Hooke case, bookkeeping both stress
and strain is redundant.

 \begin{figure}[t!]
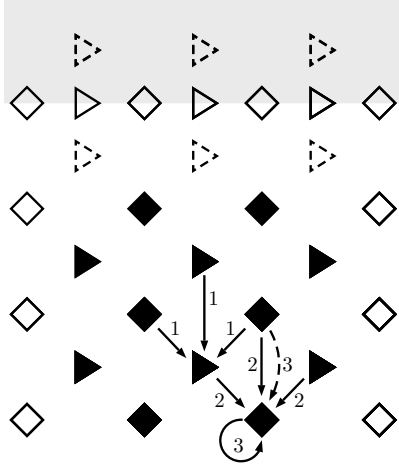
  %% .22-rol indultam, .36 a max.
%\ig{width=.22\textwidth}{1D_FT_KR_scheme_draft1_v4-kr-190419c.pdf}
\ig{width=.33\textwidth}{1D_FT_KR_scheme_draft1_v4-kr-190419c.pdf}
\caption{Visualization of the finite difference numerical scheme.
Velocity values stay at triangles, strain and stress values at
rhombuses, filled symbols denote values calculated via the scheme, while
empty ones represent initial and boundary conditions. First, new
velocities are determined from \re{Ad}, then new strains according to
\re{Bd}, and finally new stress values are obtained from \re{HCd} or
\re{Cd}, respectively. Grey indicates initial condition values \1 1
{which are typically known for a whole time interval in practice}. If
the 'grey dashed triangles' are not available then an explicit Euler
step can be used to produce the 'white dashed triangles' for starting
the scheme. }
 \label{figsema}
 \end{figure}

Rewriting the scheme for the Hooke case as
 \begin{align}  \label{Hsympl}
 &&
%\qv^{\qj+1}_{\qn} & = \qv^{\qj}_{\qn} + \qcC \f {\qDt}{\qDx}
\qv^{\qjp}_{\qnp} & = \qv^{\qjm}_{\qnp} + \qcC \f {\qDt}{\qDx}
\9 1 { \qeps^{\qj}_{\qn+1} - \qeps^{\qj}_{\qn} } ,
% \\  \label{Hooke_epsn+1}
&
\qeps^{\qj+1}_{\qn} & = \qeps^{\qj}_{\qn} + \f {\qDt}{\qDx}
%\9 1 { \qv^{\qj+1}_{\qn} - \qv^{\qj+1}_{\qn-1} },
\9 1 { \qv^{\qjp}_{\qnp} - \qv^{\qjp}_{\qnm} },
 &&
 \end{align}
we can recognize the steps of the symplectic Euler method \cite{hairer}
\1 1 {with the Hamiltonian corresponding to \m { \kin\qe + \ela\qe }}.
Now, a symplectic method is highly favourable because of its extremely
good large-time behaviour, including preservation of energy/\etc
conservation.
%\Red{Moreover, we are emphasizing that using a symplectic time stepping for a dissipative model such as the PTZ one, preserves only the total energy. This is also essential and will be discussed in detail in Section 5.}
While \re{Hsympl} coincides with the symplectic Euler
method computationally, the present interpretation of the quantities is
different, because of the space and time \Red{staggering}. One advantageous
consequence is that, due to the reflection symmetries \1 1 {see
Figure~\ref{figsema}}, our scheme
% is expected to make
 makes second order
\Red{accurate} predictions \1 1 {understood in powers of \m { \qDt } and \m {
\qDx }}, while the symplectic Euler method makes only first order
\Red{accurate} ones \cite{hairer}. Indeed, setting
 \begin{align}  \label{txv}
\qt = \qt^{\qj} ,  \qquad  \qx = \qx_{\qn}  + \ff {\qDx}{2}
 \end{align}
and assuming that
 \begin{align}  \label{vepspontos}
%\qv^{\qj}_{\qn}     & = \qv   \0 1 { \qt - \fF {\qDt}{2}, \qx } , &
\qv^{\qjm}_{\qnp}     & = \qv   \0 1 { \qt - \fF {\qDt}{2}, \qx } , &
\qeps^{\qj}_{\qn+1} & = \qeps \0 1 { \qt, \qx + \fF {\qDx}{2} } , &
\qeps^{\qj}_{\qn}   & = \qeps \0 1 { \qt, \qx - \fF {\qDx}{2} }
 \end{align}
%\Red{accurately},
% precisely,
 exactly,
the error of the prediction for \m { \qv^{\qj+1}_{\qn} } is
 \begin{align}\begin{split}  \label{vepserror}
%\qv^{\qj+1}_{\qn} - \qv \0 1 { \qt + \fF {\qDt}{2}, \qx } & =
%\qv^{\qj}_{\qn} + \qcCF \fF {\qDt}{\qDx} \9 1 { \qeps^{\qj}_{\qn+1}
\qv^{\qjp}_{\qnp} - \qv \0 1 { \qt + \fF {\qDt}{2}, \qx } & =
\qv^{\qjm}_{\qnp} + \qcCF \fF {\qDt}{\qDx} \9 1 { \qeps^{\qj}_{\qn+1}
- \qeps^{\qj}_{\qn} } - \qv \0 1 { \qt + \fF {\qDt}{2}, \qx }
 \\
& = \qv \0 1 { \qt - \fF {\qDt}{2}, \qx } - \qv \0 1 { \qt + \fF
{\qDt}{2}, \qx } + \qcCF \fF {\qDt}{\qDx} \9 2 {
\qeps \0 1 { \qt, \qx + \fF {\qDx}{2} } -
\qeps \0 1 { \qt, \qx - \fF {\qDx}{2} } }
 \\  %\nonumber
& = - \fF {\pd \qv}{\pd \qt} \0 1 { \qt, \qx } \qDt + \Ordo \0 1 {
\qDt^3 } + \qcCF \fF {\qDt}{\qDx}
\9 2 { \ff {\pd \qeps}{\pd \qx} \0 1 { \qt, \qx } \qDx +
\Ordo \0 1 { \qDx^3 } }
 \\  %\nonumber
& = \Ordo \0 1 { \qDt^3 } + \Ordo \0 1 { \qDt \qDx^2 }
 \end{split}\end{align}
after Taylor series expansion, cancellations, and the use of \re{A}.

Analogously, with
 \begin{align}  \label{txveps}
\qt = \qt^{\qj} + \ff {\qDt}{2} ,  \qquad  \qx = \qx_{\qn} ,
 \end{align}
second order \Red{accuracy} of prediction \m { \qeps^{\qj+1}_{\qn} } can be
proved.

In case of the \Red{PTZ} model, we need to discretize \re{C}. Here, both \m
{ \qsig } and its derivative, and both \m { \qeps } and its derivative,
appear. Hence, \Red{staggering} does not directly help us. This is what one can
expect for dissipative, irreversible, relaxation type, equations in
general.
 However, an interpolation-like solution is possible:
 \begin{align}  \label{Cd}
  \qalp \qsig^{\qj}_{\qn} +
 \1 1 {1 - \qalp} \qsig^{\qj+1}_{\qn} +
\qtau \f { \qsig^{\qj+1}_{\qn} - \qsig^{\qj}_{\qn} }{\qDt} & =
\qEY \9 2 {
 \qalp \qeps^{\qj}_{\qn}
+
 \1 1 {1 - \qalp} \qeps^{\qj+1}_{\qn}
} + \qEE \f { \qeps^{\qj+1}_{\qn} - \qeps^{\qj}_{\qn} }{\qDt} ,
 \end{align}
where \m { \qalp = 1/2 } is expected to provide second order \Red{accurate}
prediction, and other seminal values are \m { \qalp = 1 } \1 1 {the
explicite case, which is expected to be stiff} and \m { \qalp = 0 } \1 1
{the fully implicite case}.

For generic \m { \qalp }, \re{Cd} looks implicit. However, actually,
thermodynamics has brought in an \emph{ordinary} differential equation
type extension to the Hooke continuum, not a \emph{partial} differential
equation type one, and a linear one, in fact. Thus \re{Cd} can be
rewritten in explicit form,
 \begin{align}  \label{Cdd}
\qsig^{\qj+1}_{\qn} = \f {1}{ \1 0 {1 - \qalp} + \f {\qtau}{\qDt} }
\9 3 { \9 1 {\f {\qtau}{\qDt} - \qalp } \qsig^{\qj}_{\qn}
+ \qEY \9 2 {
 \qalp \qeps^{\qj}_{\qn}
+ 
 \1 1 {1 - \qalp} \qeps^{\qj+1}_{\qn}
} + \qEE \f { \qeps^{\qj+1}_{\qn} - \qeps^{\qj}_{\qn} }{\qDt} } ,
 \end{align}
assuming
 \begin{align}  \label{nonzero}
\1 0 {1 - \qalp} + \f {\qtau}{\qDt} \ne 0 .
 \end{align}
Second order \Red{accuracy} of \re{Cdd} for \m { \qalp = 1/2 } is then
straightforward to verify, in complete analogy to the two previous
proofs.

\section{Stability}

One may specify a space step \m { \qDx } according to the given need,
adjusted to the desirable spatial resolution. In parallel, the time step
\m { \qDt } is reasonably chosen as considerably smaller than the
involved time scales \1 1 {\eg \m { \qtau } and \m { \qtauu } of our
example system}. Now, a finite difference scheme may prove to be
unstable for the taken \m { \qDx } and \m { \qDt }, making numerical
errors \1 1 {which are generated unavoidably because of floating-point
round-off} increase essentially exponentially and ruining
%\textcolor{red}
{the} usefulness of
what we have done. Therefore, first, a stability analysis is
recommended, to explore the region of good pairs of \m { \qDx }, \m {
\qDt } for the given scheme and system.

We continue with this step for our scheme and system, performing a von
Neumann investigation \cite{charney:1950}, where the idea is similar to
the derivation of the dispersion relation. There, we study time
evolution of continuum Fourier modes \1 2 {see \re{mode}}, while here
examine whether errors, expanded in modes with \m { \e^{\ii \qk
\qx_{\qn} } } space dependence, increase or not, during an iteration by
one time step. For such linear situations as ours -- when the iteration
step means a multiplication by a matrix --, such a mode may simply
get a growth factor \m { \qxi } \1 1 {that is \m { \qk } dependent but
space independent}, in other words, the iteration matrix \1 1
{frequently called `transfer matrix'} has these modes as eigenvectors
with the corresponding eigenvalues \m { \qxi }. Then,
% \m { \9 6 { \qxi } > 1 } \1 1 {for at least one \m { \qk }}
%indicates instability \1 1 {error will exponentially increase}, and
\m { \9 6 { \qxi } < 1 } \1 1 {for all \m { \qk }} ensures stability.
Furthermore, \m { \9 6 { \qxi } = 1 } means stability if the algebraic
multiplicity of \m { \qxi } -- its multiplicity as a root of the
characteristic polynomial of the transfer matrix -- equals its geometric
multiplicity -- the number of linearly independent eigenvectors \1 1
{eigenmodes}, \ie the dimension of the eigensubspace
%, and means instability otherwise
\1 1 {\cite{elaydi}, page 186, Theorem 4.13; \cite{matolcsi:2017},
page 381, Proposition 2}.
%the proof can be found in Hungarian on page 116 in \cite{matolcsi:1995})

%\textcolor{red}
{We find important to emphasize the following. The
stability of the corresponding physical model, the \PTZ \ body is
ensured by the second law of thermodynamics. Thus asymptotic stability
of the solutions is guaranteed. The numerical method, and thus the
stability analysis must reflect the thermodynamical (physical)
requirements as well, together with the particular conditions related to
the applied discretization method. In overall, these aspects are not
independent of each other. Such a way of thinking is also emphasized in
\cite{BalKapp17}, in which a numerical method is developed for
electrodynamical problems using \Red{staggered} fields and expecting similar
properties as in our case.}

With boundary conditions specified, one can say more.\footnote{All
systems require boundary or asymptotic conditions. We also specify some
in the forthcoming section on applications.} Boundary conditions may
allow only certain combinations of \m { \e^{\ii \qk \qx_{\qn} } } as
eigenmodes of the transfer matrix. Consequently, this type of analysis
is more involved and is, therefore, usually omitted. As a general
rule-of-thumb, one can expect that \m { \9 6 { \qxi } > 1 } for some \m
{ \e^{\ii \qk \qx_{\qn} } } indicates instability also for modes obeying
the boundary conditions, while \m { \9 6 { \qxi } \le 1 } for all \m {
\e^{\ii \qk \qx_{\qn} } }
% envisions
 suggests
stability for all modes allowed by the boundary
conditions\footnote{Namely, the problem
% with
 of differing
multiplicities for \m { \9 6 { \qxi } = 1 } can be wiped out by the
boundary conditions.}.

\subsection{Hooke case}

In the Hooke case, the 'plane wave modes' for the two bookkept
quantities \m { \qv }, \m { \qeps } can, for later convenience, be
written as
 \begin{align}  \label{neumann}
%\qv^{\qj}_{\qn}    &= \ii \qAv^{\qj} \e^{\ii \qk \01{\qn+\f{1}{2}} \qDx}, &
\qv^{\qjm}_{\qnp}  &= \ii \qAv^{\qj} \e^{\ii \qk \01{\qn+\f{1}{2}} \qDx}, &
\qeps^{\qj}_{\qn}  &= \qAeps^{\qj}   \e^{\ii \qk \qn \qDx}, &
%\qsig^{\qj}_{\qn} &= \qAsig^{\qj}   \e^{\ii \qk \qn \qDx}, &
\qk\qDx \in [ 0, 2\pi ) ,
 \end{align}
the condition on \m { \qk } related to that \m { \qk } outside such a
'Brillouin zone' makes the description redundant.

Realizing the iteration steps \re{Hsympl} as matrix products leads, for
the amplitudes introduced in \re{neumann}, to
 \begin{align}  \label{matrix_Hooke}
  \begin{split}
\begin{pmatrix} \qAv^{\qj+1} \\ \qAeps^{\qj+1} \end{pmatrix} & =
%\underbrace{
\begin{pmatrix} 1 & 0 \\  -2 \f{\qDt}{\qDx} \qS & 1 \end{pmatrix}
%}_{=:\qqT_{{\rm{H}},\qeps}}
\cdot \begin{pmatrix} \qAv^{\qj+1} \\ \qAeps^{\qj} \end{pmatrix} =
\begin{pmatrix} 1 & 0 \\ -2\f{\qDt}{\qDx}\qS & 1 \end{pmatrix} \cdot
%\underbrace{
\begin{pmatrix} 1 & 2 \qc^2 \f{\qDt}{\qDx} \qS \\ 0 & 1 \end{pmatrix}
%}_{=:\qqT_{{\rm{H}},\qv}}
\cdot \begin{pmatrix} \qAv^{\qj} \\ \qAeps^{\qj} \end{pmatrix}
% \nonumber
 \\
% & \\ \nonumber
& =
%\underbrace{
\begin{pmatrix} 1 & 2 \qc^2 \f{\qDt}{\qDx} \qS \\ -2 \f{\qDt}{\qDx}
\qS & 1 - 4 \qc^2 \f{\qDt^2}{\qDx^2} \qS^2 \end{pmatrix}
%}_{=:\qqT_{{\rm{H}}}}
\cdot \begin{pmatrix} \qAv^{\qj} \\ \qAeps^{\qj} \end{pmatrix} \equiv
\qqh \qqT \cdot \begin{pmatrix} \qAv^{\qj} \\ \qAeps^{\qj} \end{pmatrix}
%\qquad\quad \text{with} \quad \qS = \qSS , \quad 0 \le \qS \le 1 .
 \end{split}
 \end{align}
with
 \begin{align}  \label{S}
\qS = \qSS , \qquad 0 \le \qS \le 1 .
 \end{align}
For space dependences \re{neumann},
 \begin{align} \label{growth_factor}
\qv^{\qjp}_{\qnp} = \qxi \qv^{\qjm}_{\qnp} ,  \quad
\qeps^{\qj+1}_{\qn} = \qxi \qeps^{\qj}_{\qn}  \qquad \text{lead to} \qquad
\qAv^{\qj+1} = \qxi \qAv^{\qj} , \quad \qAeps^{\qj+1} = \qxi \qAeps^{\qj} ,
 \end{align}
in other words, to the eigenvalue problem
 \begin{align} \label{eigenvalue}
\qqh\qqT \qy = \qxi \qy  \qquad \text{with} \qquad
\qy = \begin{pmatrix} \qAv^{\qj} \\ \qAeps^{\qj} \end{pmatrix} .
 \end{align}

Let us introduce the notation
 \begin{align}  \label{Cou}
\qC = \qc \f{\qDt}{\qDx}
 \end{align}
for the Courant number of our scheme for the Hooke system.
%Then, c
Comparing the characteristic polynomial of \m{\qqT},
 \begin{align}  \label{P_2}
\qP
\0 1 {\qxi} =
%\qxi^2 - 2 \9 2 { 1 - 2 \qC^2 \qS^2 } \qxi + 1
\qxi^2 + \9 1 { 4 \qC^2 \qS^2 - 2 } \qxi + 1
 \end{align}
with its form written via its roots,
 \begin{align}  \label{hrf}
\9 1 { \qxi - \qxii } \9 1 { \qxi - \qxiii } = \qxi^2 - \9 1 { \qxii +
\qxiii } \qxi + \qxii \qxiii ,
 \end{align}
reveals, on one side, that, in order to have both \m { \9 6 { \qxii }
\le 1 } and \m { \9 6 { \qxii } \le 1 }, both
magnitudes have to be 1 \1 1 {since their product is 1}, which, on the
other side, also implies
 \begin{align}  \label{vi1}
4 \qC^2 \qS^2 - 2 = - \qxii - \qxiii \le \9 6 { \qxii } + \9 6 { \qxiii
} \le 2  \qquad  \Longrightarrow  \qquad  \qC^2 \qS^2 \le 1 ,
\quad \qC \qS \le 1
 \end{align}
as both \m { \qC } and \m { \qS } are non-negative.

If \m { \qC \qS < 1 } then the two roots,
 \begin{align} \label{roots}
\qxiiii = 1 - 2 \qC^2 \qS^2 \pm
\sqrt{ 4 \qC^2 \qS^2 \9 1 { \qC^2 \qS^2 - 1 } } ,
 \end{align}
are complex, with unit modulus, and are the complex conjugate of one
another. Especially simple -- and principally distinguished, as we see
in the next sections -- is the case \m { \qC = 1 }: then
 \begin{align}  \label{hc1}
\qxiiii = \e^{ \pm \ii \qk \qDx } ,
 \end{align}
with the remarkable property that \m { \arg \qxiiii } are linearly
depending on \m { \qk } -- so to say, both branches of the discrete
dispersion relation are linear.

In parallel, if \m { \qC \qS = 1 } then the two roots coincide, \m {
\qxiiii = -1 }. The algebraic multiplicity 2 is accompanied with
geometric multiplicity 1: only the multiples of
 \begin{align}  \label{eig}
\qy = \begin{pmatrix} \qc \\ - 1 \end{pmatrix}
 \end{align}
are eigenvectors. If \m { \qC = 1 } then this affects only one mode, \m
{ \qS = 1 }, \m { \qk = \f {\pi}{\qk} }, and if that mode is prohibited
by the boundary conditions then the choice \m { \qC = 1 } ensures a
stable scheme.

With \m { \qC > 1 }, \m { \qC \qS \le 1 } would be violated by a whole
interval for \m { \qk } \1 2 {recall \re{S}}, which may not be cured by
boundary conditions so the best candidate \1 1 {largest \m { \qDt } for
a fixed \m { \qDx }, or the smallest possible \m { \qDx } for fixed \m {
\qDt }} to have stability is \m { \qC = 1 }.

\subsection{\PTZ\ case}

For the \Red{PTZ} system, the von Neumann stability analysis of our
discretization studies the modes
 \begin{align}  \label{neumannn}
%\qv^{\qj}_{\qn}   &= \ii \qAv^{\qj} \e^{\ii \qk \01{\qn+\f{1}{2}} \qDx}, &
\qv^{\qjm}_{\qnp} &= \ii \qAv^{\qj} \e^{\ii \qk \01{\qn+\f{1}{2}} \qDx}, &
\qeps^{\qj}_{\qn} &= \qAeps^{\qj}   \e^{\ii \qk \qn \qDx}, &
\qsig^{\qj}_{\qn} &= \qAsig^{\qj}   \e^{\ii \qk \qn \qDx},
 \end{align}
on which iteration via
 \begin{align}  \label{pv}
%\qv^{\qj+1}_{\qn} & = \qv^{\qj}_{\qn} + \f {1}{\qrho} \f {\qDt}{\qDx}
\qv^{\qjp}_{\qnp} & = \qv^{\qjm}_{\qnp} + \f {1}{\qrho} \f {\qDt}{\qDx}
\9 1 { \qsig^{\qj}_{\qn+1} - \qsig^{\qj}_{\qn} } ,
% \\  \label{pe}
\qquad\qquad\quad
\qeps^{\qj+1}_{\qn} = \qeps^{\qj}_{\qn} + \f {\qDt}{\qDx}
%\9 1 { \qv^{\qj+1}_{\qn} - \qv^{\qj+1}_{\qn-1} },
\9 1 { \qv^{\qjp}_{\qnp} - \qv^{\qjp}_{\qnm} },
 \\  \label{ps}
\qsig^{\qj+1}_{\qn} & = \f {1}{ \1 0 {1 - \qalp} + \f {\qtau}{\qDt} }
\9 3 { \9 1 {\f {\qtau}{\qDt} - \qalp } \qsig^{\qj}_{\qn}
+ \qEY \9 2 { \qalp \qeps^{\qj}_{\qn} + \1 1 {1 - \qalp} \qeps^{\qj+1}_{\qn}
} + \qEE \f { \qeps^{\qj+1}_{\qn} - \qeps^{\qj}_{\qn} }{\qDt} } ,
 \end{align}
gives
 \begin{align}  \label{TT}  \begin{split}
\begin{pmatrix} \qAv^{\qj+1} \\ \qAeps^{\qj+1} \\ \qAsig^{\qj+1} \end{pmatrix}
& =
%\underbrace{\underbrace{
\begin{pmatrix} 1 & 0 & 0 \\ 0 & 1 & 0 \\ 0 & \f{ \qEY \0 1 {1 - \qalp} +
\f{\qEE}{\qDt} }{ \0 1 {1 - \qalp} + \f{\qtau}{\qDt} } & \f{ \f{\qtau}{\qDt}
- \qalp }{ \0 1 {1 - \qalp} + \f{\qtau}{\qDt} } \end{pmatrix}
%}_{=:\qqT_{{{---PT---}},\qsig,1}}
\cdot
%\underbrace{\underbrace{
   \begin{pmatrix}
   1 & 0 & 0\\
   -2\f{\qDt}{\qDx}\qS & 1 & 0\\
   0 & 0 & 1
   \end{pmatrix}
%}_{=:\qqT_{{{---PT---}},\qeps}}
\cdot
%\underbrace{\underbrace{
\begin{pmatrix} 1 & 0 & 2 \f{\qDt}{\qrho\qDx} \qS \\
0 & 1 & 0 \\ 0 & 0 & 1 \end{pmatrix}
%}_{=:\qqT_{{{---PT---}},\qv}}
\cdot
   \begin{pmatrix}
   \qAv^{\qj} \\ \qAeps^{\qj} \\ \qAsig^{\qj}
   \end{pmatrix}
%}_{=  \begin{pmatrix} \qAv^{\qj+1} & \qAeps^{\qj} & \qAsig^{\qj}
%   \end{pmatrix}^{{{\rm T}}}
%}}_{= \begin{pmatrix} \qAv^{\qj+1} & \qAeps^{\qj+1} & \qAsig^{\qj}
%  \end{pmatrix}^{{{\rm T}}}}
%}_{= \begin{pmatrix} \qAv^{\qj+1} & \qAeps^{\qj+1} & {\qAAsig{1}^{\qj+1}}
% \end{pmatrix}^{{{\rm T}}}}
 \\
& \quad +
%\underbrace{\underbrace{
\begin{pmatrix} 0 & 0 & 0 \\ 0 & 0 & 0 \\ 0 &
\f{\qEY\qalp - \f{\qEE}{\qDt} }{ \0 1 {1 - \qalp} + \f{\qtau}{\qDt} } & 0
\end{pmatrix}
%}_{=:\qqT_{{{---PT---}},\qsig,0}}
\cdot \begin{pmatrix} \qAv^{\qj} \\ \qAeps^{\qj} \\ \qAsig^{\qj} \end{pmatrix}
%}_{=\begin{pmatrix} 0 & 0 & {\qAAsig{0}^{\qj+1}} \end{pmatrix}^{{{\rm T}}}}.
\equiv \qqTT
\begin{pmatrix} \qAv^{\qj} \\ \qAeps^{\qj} \\ \qAsig^{\qj} \end{pmatrix}
 \end{split}  \end{align}
with
 \begin{align}  \label{TTT}  \begin{split}
%\qqT_{{{---PT---}}} &=\qqT_{{{---PT---}},\qsig,1}\cdot\qqT_{{{---PT---}},\qeps}\cdot\qqT_{{{---PT---}},\qv}+\qqT_{{{---PT---}},\qsig,0}\\
\qqTT = \begin{pmatrix}
1 & 0 & 2 \f{\qDt}{\qrho\qDx} \qS \\
-2 \f{\qDt}{\qDx} \qS & 1 & -4 \f{\qDt^2}{\qrho\qDx^2} \qS^2 \\
-2 \f{ \qEY \0 1 {1 - \qalp} + \f{\qEE}{\qDt} }{ \0 1 {1 - \qalp} +
\f{\qtau}{\qDt} } \cdot \f{\qDt}{\qDx} \qS & \f{\qEY}{ \0 1 {1 - \qalp} +
\f{\qtau}{\qDt} } & \f{ \f{\qtau}{\qDt} - \qalp }{ \0 1 {1 - \qalp} +
\f{\qtau}{\qDt} } -4 \f{ \qEY \0 1 {1 - \qalp} + \f{\qEE}{\qDt} }{\0 1 {1 -
\qalp} + \f{\qtau}{\qDt} } \cdot \f{\qDt^2}{\qrho\qDx^2} \qS^2 \end{pmatrix} .
 \end{split}  \end{align}

The characteristic polynomial is now
 \begin{align} \label{polxi}
\qPP \0 1 {\qxi} = \qai{3} \qxi^3 + \qai{2} \qxi^2 + \qai{1} \qxi + \qai{0} ,
 \end{align}
%with
 \begin{align}  \label{a0123}
\qai{0} & =
\f{ \qalp - \f{\qtau}{\qDt} }{ \0 1 {1 - \qalp} + \f{\qtau}{\qDt} } ,
 &
\qai{1} & =
3 - \f { 2 - 4 \0 1 {\qalp - \f{\qtauu}{\qDt}} \qC^2 \qS^2 }
{ \0 1 {1 - \qalp} + \f{\qtau}{\qDt} } ,
 &
\qai{2} & =
-3 + \f { 1 + 4 \9 2 { \0 1 {1 - \qalp} + \f{\qtauu}{\qDt} }
\qC^2 \qS^2 }{ \0 1 {1 - \qalp} + \f{\qtau}{\qDt} } ,
 &
\qai{3} & = 1 .
 \end{align}

Three roots are considerably more difficult to
% \Red{analyze directly}.
%% !!! twice as many Google results
%% for 'directly analyze' than for Grammarly's 'analyze directly'
%% British English also OK: https://www.mdpi.com/authors/english-editing
 directly analyse.
One
alternative is to use Jury's criteria \cite{jury:1974} for whether the
roots are within the unit circle of the complex plane, and another
possibility is to apply the M\"obius transformation \m{\qxi =
\f{\qeta+1}{\qeta-1}} on \re{polxi} and utilize the Routh--Hurwitz
criteria whether the mapped roots are within the left half plane. The
two approaches provide the same result. Nevertheless, one criterion
provided by one of these two methods may not directly be one criterion
of the other method. It is only the combined result \1 1 {the
intersection of the conditions} that agrees. Accordingly, it can be
beneficial to perform both investigations because a simple condition
provided by one of the routes may be labouring to recognize as
% \Red{a}  %% !!! overruling Grammarly
consequence of the conditions directly offered by the other route.

Jury's criteria, for our case, are as follows. First,
\m{\qPP \0 1 {
%\qxi = 
1} > 0} gives
 \begin{align}  \label{P(1)}
\f{ 4 \qC^2 \qS^2 }{ \0 1 {1 - \qalp} + \f{\qtau}{\qDt} } > 0,
 \qquad  \Longleftrightarrow  \qquad
\0 1 {1-\qalp} + \f{\qtau}{\qDt} > 0 .
 \end{align}
Second, \m { \0 1 {-1}^3 \qPP \0 1 {
% \qxi =
-1 } > 0 } yields
 \begin{align}
8 - 8 \f{ \f{1}{2} + \9 2 { \0 1 {\f{1}{2} - \qalp} + \f{\qtauu}{\qDt} }
\qC^2 \qS^2 }{ \0 1 {1 - \qalp} + \f{\qtau}{\qDt} } > 0
 \end{align}
which, in light of \re{P(1)}, reduces to
 \begin{align}  \label{bb}
\0 1 {\fFF{1}{2} - \qalp} + \fFF{\qtau}{\qDt} >
\9 2 { \0 1 {\fFF{1}{2} - \qalp} + \fFF{\qtauu}{\qDt} } \qC^2 \qS^2 .
 \end{align}
Third, the matrices 
 \m{ \9 1 { \begin{smallmatrix} \qai{3} & \qai{2} \\ 0 & \qai{3}
 \end{smallmatrix} } \pm \9 1 { \begin{smallmatrix} 0 & \qai{0} \\
 \qai{0} & \qai{1} \end{smallmatrix} } }
have to be positive innerwise\footnote{Following Jury \cite{jury:1974},
a matrix is \emph{positive innerwise} if the determinant of the matrix
and its all inners are positive. Here, \emph{inner} \m{\qinner_{\qm-2}}
of an \m{\qm \times \qm} matrix is formed by deleting its first and
\m{\qm}th rows and columns, inner \m{\qinner_{\qm-4}} is the inner of
\m{\qinner_{\qm-2}}, and the procedure is continued until
\m{\qinner_{1}} or \m{\qinner_{2}} is reached. Inners enter the picture
only for \m{\qm\ge3} so, in our case, only positive definiteness of the
matrices themselves is to be ensured.}. The `\m{+}' branch leads to
% \m{ \begin{pmatrix}  \qai{3} & \qai{2}+\qai{0}\\  \qai{0} & \qai{3}+\qai{1} \end{pmatrix}}
%requires
% \begin{align}
% \f{2\02{2\qtau+\01{1-2\qalp}\qDt}\qrho\qDx^2-4\01{\qEE-\qtau\qEY}\qDt^2\qS^2}{\qrho\qDt\qDx^2\02{\01{1-\qalp}+\f{\qtau}{\qDt}}^2}>0,
% \end{align}
% so
 \begin{align}  \label{bB}
%\02{2\qtau+\01{1-2\qalp}\qDt}\qrho\qDx^2-2\01{\qEE-\qtau\qEY}\qDt^2\qS^2>0.
\0 1 {\fFF{1}{2} - \qalp} + \fFF{\qtau}{\qDt} >
\fFF {\qtauu - \qtau}{\qDt} \qC^2 \qS^2 ,
 \end{align}
which is weaker than \re{bb}, because there the \rhs\ is larger by \m {
\0 2 { \0 1 {1 - \qalp} + \f{\qtau}{\qDt} } \qC^2 \qS^2 } \1 2 {and \cf
\re{P(1)}}.
 Meanwhile, the `\m{-}' branch induces condition
% \m{ \begin{pmatrix}  \qai{3} & \qai{2}-\qai{0}\\  -\qai{0} & \qai{3}-\qai{1} \end{pmatrix}}
% \begin{align}
% \f{4\01{\qEE-\qtau\qEY}\qDt\qS^2}{\qrho\qDx^2\02{\01{1-\qalp}+\f{\qtau}{\qDt}}^2}>0,
% \end{align}
% so if \m{\qS\neq 0} then
 \begin{align}  \label{tt}
\qtauu > \qtau ,  % \qEE-\qtau\qEY>0,
 \end{align}
which we have already met in \re{tauhat}, as the thermodynamical
requirement \re{ineq} at the continuum level, and which also induces,
via \re{bB},
 \begin{align}  \label{bbb}
\0 1 {\fFF{1}{2} - \qalp} + \fFF{\qtau}{\qDt} > 0 ,
 \end{align}
which is stronger than \re{P(1)}. This also allows to rearrange \re{bb}
and exploit it as
 \begin{align}  \label{Bb}
\qC^2 \qS^2 < \f { \0 1 {\fFF{1}{2} - \qalp} + \fFF{\qtau}{\qDt} }
{ \0 1 {\fFF{1}{2} - \qalp} + \fFF{\qtauu}{\qDt} } < 1
 \qquad  \text{for all} \quad 0 \le \qS \le 1 \quad \Longrightarrow \quad
\qC^2 < \f { \0 1 {\fFF{1}{2} - \qalp} + \fFF{\qtau}{\qDt} }
{ \0 1 {\fFF{1}{2} - \qalp} + \fFF{\qtauu}{\qDt} } < 1 .
 \end{align}
Conditions \re{tt}--\re{Bb} summarize the obtained stability
requirements; the first referring to the constants of the continuum
model only, the second relating \m { \qalp } and \m { \qDt } of the
scheme, and the third limiting \m { \qDx } \1 1 {through \m { \qC }} in
light of \m { \qalp } and \m { \qDt }.

If, instead of Jury's criteria, one follows the Routh--Hurwitz path, on
the M\"obius transformed polynomial
 \begin{align}  \label{pol}
\qQQ \0 1 {\qeta} = \0 1 {\qeta -1}^3 \qPP \0 1 { \f{\qeta + 1}{\qeta - 1} }
= \qa{3} \qeta^3 + \qa{2} \qeta^2 +\qa{1} \qeta + \qa{0} ,
 \end{align}
 \begin{align}  \label{RHb}
\qa{0} & = 8 \qrho \qDx^2 \9 3 { \0 2 { \0 1 {\fFF{1}{2} - \qalp} +
\fFF{\qtau}{\qDt} } - \0 2 { \0 1 {\fFF{1}{2} - \qalp} +
\fFF{\qtauu}{\qDt} } \qC^2 \qS^2 } ,
 &
\qa{1} & = 4 \qrho \qDx^2 \0 1 {1 - \qC^2 \qS^2} \qDt ,
 \\
\qa{2} & = 8 \0 2 { \0 1 {\fFF{1}{2} - \qalp} +
\fFF{\qtau}{\qDt} } \qEY \qDt^2 \qS^2 ,
 &
\qa{3} & = 4\qEY\qDt^3\qS^2 ,
  \end{align}
then, having \m{ \qa{3} > 0 }, roots lie in the left half plane if all
corner subdeterminants of
% \begin{align}  \label{RHm}
 \m {
\9 1 { \begin{smallmatrix} \qa{2} & \qa{0} & 0 \\ \qa{3} & \qa{1} & 0 \\
0 & \qa{2} & \qa{0} \end{smallmatrix} }
% \end{align}
 }
are positive, \ie \m{ \qa{2} > 0}, \m{ \qa{1} \qa{2} - \qa{0} \qa{3} > 0
} and \m{ \qa{0} \0 1 {\qa{1} \qa{2} - \qa{0} \qa{3} } > 0 } \1 1
{hence, \m { \qa{0} > 0 }} are needed.
% to be satisfied.
As expected, these conditions prove to be equivalent to the ones
obtained via Jury's criteria -- we omit the details to avoid redundant
repetition.

\subsubsection{Kelvin--Voigt model}

Although the focus of the present paper is on the hyperbolic-like case
corresponding to \m { \qtau > 0 }, the above calculations are valid for
\m { \qtau = 0 }, the Kelvin--Voigt subfamily as well. As a brief
analysis of this case, \re{tt} is trivially satisfied with \m { \qtauu >
0 }. \re{bbb} gives the nontrivial condition \m { \qalp < \F {1}{2}
}.\footnote{Together with boundary conditions, this may be weakened to
\m { \qalp \le \F {1}{2} }%
%\ but still may forecast stiffness%
 .} Finally, \re{Bb} gives
 \begin{align}  \label{KV}
\0 1 {\fFF{1}{2} - \qalp} \qDt^2 + \qtauu \qDt <
\f { \fFF{1}{2} - \qalp }{ \qc^2 } \qDx^2 ,
 \end{align}
which looks like some mixture of a stability condition for a scheme for
a parabolic problem like Fourier heat conduction and of a condition for
a simple reversible wave propagation.

\subsubsection{Beyond Kelvin--Voigt}

When \m { \qtau > 0 } then
 \begin{align}  \label{Couu}
\qCC = \qcc \f{\qDt}{\qDx} > \qC
 \end{align}
\1 2 {recall \re{cc}} becomes important.

The most interesting case is
 \underline{%
\m { \qalp = \F{1}{2} }%
 }%
, where the scheme gives second order \Red{accurate} predictions: \re{bbb}
holds trivially, and \re{Bb} can be rewritten as
 \begin{align}  \label{qCC}
\qCC < 1 .
 \end{align}
With boundary conditions also present, we may extend this condition to
 \begin{align}  \label{qCCC}
\qCC \le 1 .
 \end{align}

Considering the two other potentially interesting cases as well: If
 \underline{%
\m { \qalp = 1 }%
 }
then \re{bbb} induces \m { \qDt < 2 \qtau }, which is not a harsh
requirement since the time step must usually be much smaller then the
time scales of the system in order to obtain a physically acceptable
numerical solution. In parallel, \m { \qCC } is limited from above by a
number smaller than 1. On the other side, when
 \underline{%
\m { \qalp = 0 }%
 }
then \re{bbb} is automatically true again, and now \m { \qCC } is
limited from above by a number larger than 1. Since we may need \m {
\qDt \ll \qtau } for a satisfactory solution, this \m { \Ordo \9 1 { \f
{\qDt}{\qtau} } } gain over 1 is not considerable.

\subsubsection{Hooke case}

It is worth looking back to the Hooke limit of \re{Bb}: \m { \qtau =
\qtauu = 0 } \1 1 {with whatever \m { \qalp }} tells \m { \qC < 1 }. One
can see that the \m { \9 6 { \qxi } < 1 } stability requirement gives
conservative results and does not tell us how far the obtained
inequalities are from equalities.

\section{Numerical results}

The calculations communicated here are carried out with zero \m { \qv,
\qeps, \qsig } as initial conditions, and with stress boundary
conditions: on one end of the sample, a cosine shaped pulse is applied,
while the other end is free \1 1 {stress is zero}. With \m { \qtaub }
denoting the temporal width of the pulse, the excitation is,
hence,
 \begin{align}  \label{pulse}
\qsig \0 1 { \qt, 0 } = \begin{cases}
\f {\qsigb}{2} \9 2 { 1 - \cos \0 1 { 2 \pi \f {\qt}{\qtaub} } }
& \text{if} \quad 0 \le \qt \le \qtaub ,
 \\
0 & \text{otherwise}.
\end{cases}
 \end{align}
Temperature is calculated
% according to
 via
the discretized form of \re{Tdot}, with the natural choice that
temperature values reside at the same place as stress and strain but
half-shifted in time \1 1 {\m { \qT_\qn^{\qjm} } at \m { \qt^{\qj} - \ff
{\qDt}{2}, \; \qx_{\qn} }}.

When plotting, say, elastic energy of the whole sample at time \m {
\qt^\qj }, a simple \m { \f {\qEY}{2}\sum_{\qn
%=0}^\qN
 }
\0 1 { \qeps_\qn^\qj }^2 \qDx } type sum is used, with two adjustments.
First, terms living at the outer endpoint of an outermost space cell,
such as \m { \0 1 { \qeps_0^\qj }^2 } and \m { \0 1 { \qeps_\qN^\qj }^2
}, are counted with weight \m { \f {1}{2} }. Second, kinetic energy and
thermal energy, both being based on quantities half-shifted in time, are
calculated as a time average, their value at \m { \qt^\qj } taken as the
average of their value at \m { \qt^{\qj} - \ff {\qDt}{2} } and \m {
\qt^{\qj} + \ff {\qDt}{2} }.

The numerical calculations are performed for dimensionless quantities.
For making the quantities dimensionless, the following units are used:
the length of the sample \m { \qX }, \m { \qc } \1 1 {so a Hookean wave
arrives at the other end during unit time}, \m { \qEY }, \m { \qsigb },
and
%, for temperature,
\m { \qcp }. Accordingly, dimensionless position, time, velocity,
stress, strain, energy, temperature, and wave number are defined as
 \begin{align}  \label{dimless}
\nd{\qx} & = \f {1}{\qX} \qx ,  &  \nd{\qt} & = \f {\qc}{\qX} \qt ,  & 
\nd{\qv} & = \f {1}{\qc} \qv ,  &  \nd{\qsig} & = \f {1}{\qsigb} \qsig
,  &  \nd{\qeps} & = \f {\qEY}{\qsigb} \qeps ,  &  \nd{\qe} & = \f
{\qEY^2}{\qc^2 \qsigb^2} \qe ,  &  \nd{\qT} & = \f
{\qEY^2 \qcp}{\qc^2 \qsigb^2} \qT ,  &  \nd{\qk} = \qX \qk .
 \end{align}
%% c^2 = EY/rho   rho = EY/c^2
%% rho cp T/t = sig^2/Ehat = sigb^2/(EY*tu)
%% T = (1/cp) sigb^2/{EY rho} = sigb^2 /{EY rho} /cp =
%%   = sigb^2 / (EY^2/c^2) /cp = sigb^2 c^2 / EY^2 / cp =
%%   Pa^2 m^2/s^2 / Pa^2 / (J/kg K) = m^2 / s^2 / (m^2/s^2/K) = K
% On this time scale, 
%Henceforth, with respect to this time unit, \m { 0.2 } is used for \m {
%\qtaub }, \m { 1.25 } for \m { \qtau }, and \m { 5 } for \m { \qtauu },
%implying \m { \qcc = 2 \qc \equiv 2 }.
The results are presented for dimensionless time constants
 \begin{align}  \label{dimlesss}
 &&
\qndtaub & = 0.2 ,  &  \nd{\qtau} & = 1.25 ,  &  \nd{\qtauu} & = 5 ,
 &&
 \end{align}
the latter two implying \m { \qcc / \qc = 2 }.

 \begin{figure}[H]
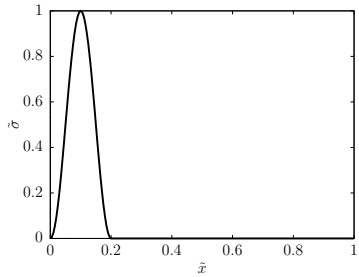
  %% width=.32\textwidth
\ig{height=\qFIG}{N200-tau_b0_2-Hooke-S1-End2-First1.pdf}
 \caption{Snapshot of the shape of the fully born stress pulse near the
left end of the sample.}
 \label{figpulse}
 \end{figure}

\subsection{Hookean wave propagation}

\begin{figure}[H]
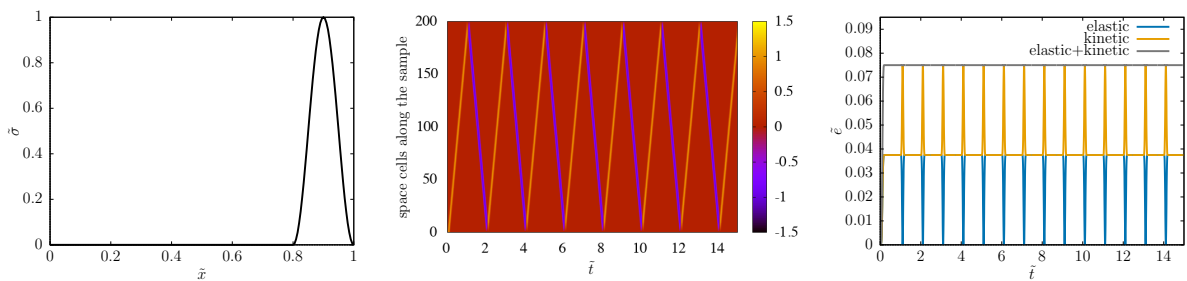

\iggg{height=\qFIG}  %\iggg{width=.32\textwidth}  height=20ex
{N200-tau_b0_2-Hooke-taumax15-S1-End2-Last1.pdf}
{N200-tau_b0_2-Hooke-taumax15-S1-End2-ST2.pdf}
{N200-tau_b0_2-Hooke-taumax15-S1-End2-Energy1.pdf}
 \caption{\emph{Left:} snapshot of the stress pulse right before its 15th
bouncing back from the boundary. \emph{Middle:} spacetime picture of the
wave propagation. Bouncing back from free ends makes stress change sign.
\emph{Right:} elastic energy, kinetic energy, and their sum as functions
of time. Calculation done with \m { \qN = 200 } space cells and \m { \qC
= 1 }.}
 \label{figH}
 \end{figure}

For the Hooke system, our scheme is symplectic, with very reliable
long-time behaviour. This is well visible in Figure~\ref{figH}: the
shape is nicely preserved, no numerical artefacts are visible in the
spacetime picture, and the sum of elastic and kinetic energy is
conserved.

\subsection{\PTZ\ wave propagation}

\begin{figure}[H]
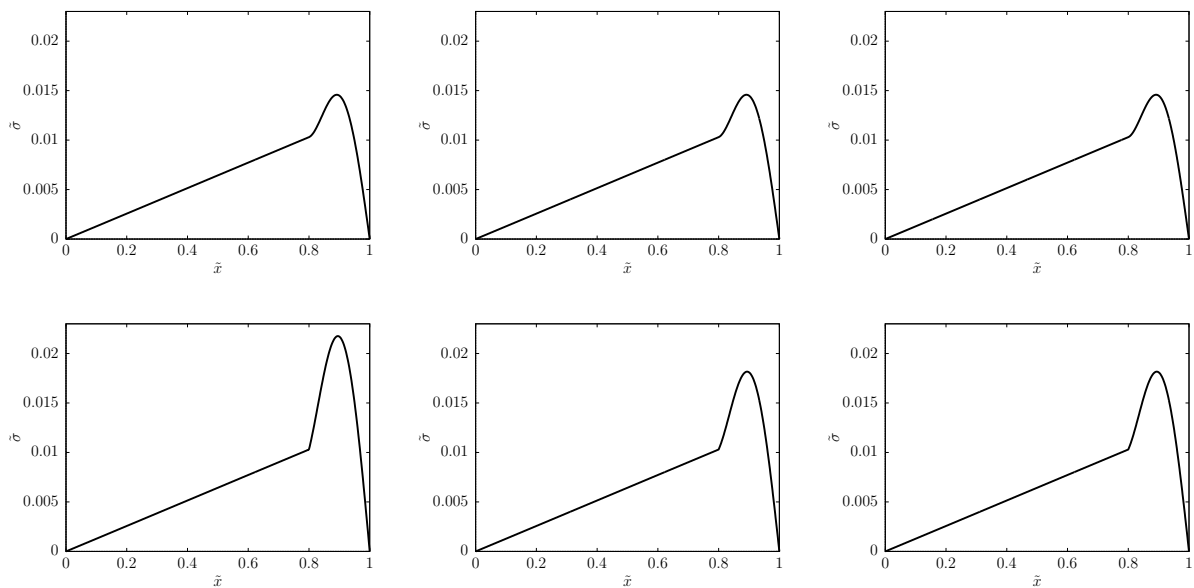
  %% width=.32\textwidth
\iggg{height=\qFIG}
{N400-tau_b0_2-a0_5-tau_C5-RC0_25-taumax3_6-S1-End2-Y0_023.pdf}
{N800-tau_b0_2-a0_5-tau_C5-RC0_25-taumax3_6-S1-End2-Y0_023.pdf}
{N1600-tau_b0_2-a0_5-tau_C5-RC0_25-taumax3_6-S1-End2-Y0_023.pdf}
 \par\vspace{2.5 ex}
\iggg{height=\qFIG}
{N400-tau_b0_2-a0-tau_C5-RC0_25-taumax3_6-S1-End2-Y0_023.pdf}
{N800-tau_b0_2-a0-tau_C5-RC0_25-taumax3_6-S1-End2-Y0_023.pdf}
{N1600-tau_b0_2-a0-tau_C5-RC0_25-taumax3_6-S1-End2-Y0_023.pdf}
 \caption{\emph{Upper row:} \m { \qalp = 1/2 }, \emph{Lower row:}
\m { \qalp = 0 } calculation of the stress signal when it starts its 7th
bouncing, with \m { \qCC = 1 }. \emph{From left to right:}
\m { N = 400, \, 800, \, 1600 } space cells.}
 \label{figPTZ}
 \end{figure}

For the \Red{PTZ} system, we find that the principally optimal choice of \m
{ \qalp = 1/2 } does outperform \m { \qalp = 0 } \1 1 {with \m { \qCC =
1 }}. Figure~\ref{figPTZ} shows such a comparison: \m { \qalp = 1/2 }
produces a reliable signal shape quite independently of space
resolution, while \m { \qalp = 0 } needs more than \m { \qN = 1000 }
space cells to reach the same reliability.
%right before the 7th bouncing

\begin{figure}[H]
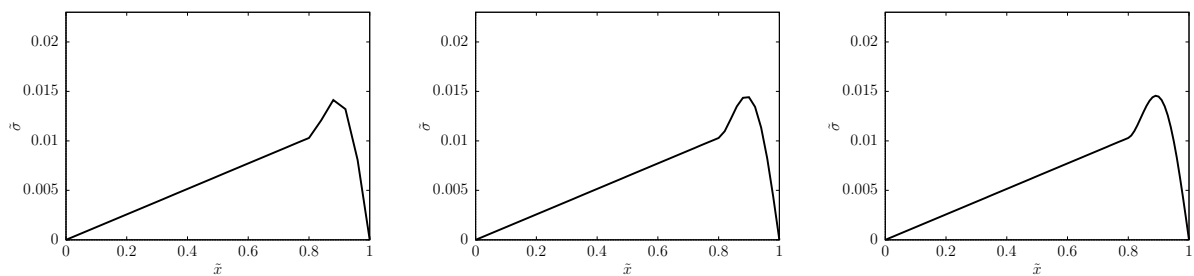

\iggg{height=\qFIG}  %% \iggg{width=.32\textwidth}  %\igggg{width=.24\textwidth}
{N25-tau_b0_2-a0_5-tau_C5-RC0_25-taumax3_6-S1-End2-Y0_023.pdf}
{N50-tau_b0_2-a0_5-tau_C5-RC0_25-taumax3_6-S1-End2-Y0_023.pdf}
{N100-tau_b0_2-a0_5-tau_C5-RC0_25-taumax3_6-S1-End2-Y0_023.pdf}
 \caption{The same \m { \qalp = 1/2 } prediction with \m { N = 25, \,
50, \, 100 } space cells, from left to right, respectively.}
 \label{figPTZZ}
 \end{figure}

Actually, \m { \qalp = 1/2 } offers that realibility already at \m { \qN
= 50 }, and even \m { \qN = 25 } `does a decent job', as depicted in
Figure~\ref{figPTZZ}.

 \begin{figure}[H]
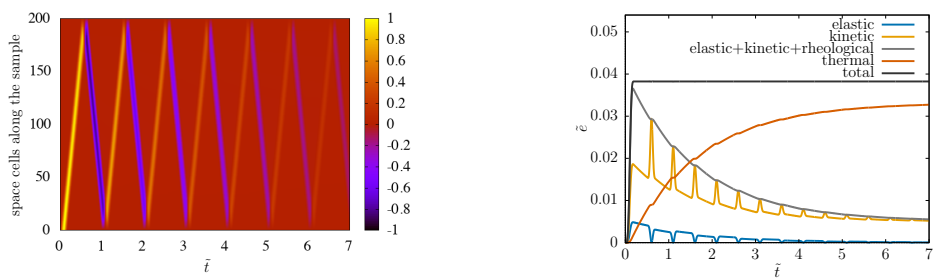
  %% height=20ex
\igg{height=\qFIG}
{N200-tau_b0_2-a0_5-tau_C5-RC0_25-taumax7-S1-End2-ST2.pdf}
{N200-tau_b0_2-a0_5-tau_C5-RC0_25-taumax7-S1-End2-Energy2.pdf}
 \caption{\m { \qalp = 1/2 }, \m { \qCC = 1 } spacetime picture and
energy conservation, \m { N = 200 }.}
 \label{figPTZZZ}
 \end{figure}

With \m { \qalp = 1/2 }, the spacetime picture and total energy
conservation are not less satisfactory, as visible in
Figure~\ref{figPTZZZ}.

The physical explanation of the signal shape  \1 1
{Figures~\ref{figPTZ}--\ref{figPTZZ}} is that the fastest modes
propagate with speed \m { \qcc } \1 1 {recall Section~2}, transporting
the front of the signal, while slow modes travel with \m { \qc < \qcc },
gradually falling behind, and forming a little-by-little thickening
tail.

In parallel, the spacetime picture shows that this tail effect is less
relevant than the overall decrease of the signal, due to dissipation.

Finally, concerning the energy results, the remarkable fact is that all
ingredients \m { \qv, \qeps, \qsig, \qT } are calculated via discretized
time integration, therefore, total energy conservation is not built-in
but is a test of the quality of the whole numerical approach.
The observed good energy conservation behaviour would deserve deeper
analysis in the future, possibly analogously to \cite{gayos18}. 
% \Red{That is, even for the dissipative PTZ model, the total energy is highly favorable to be conserved.}

\section{Dissipation error and dispersion error}

\subsection{Hooke case}

\begin{figure}[H]
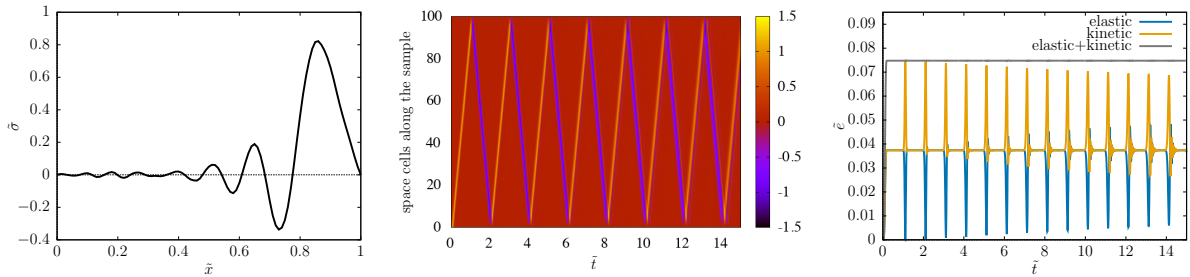
  %% height=20ex
\iggg{height=\qFIG}
{N100-tau_b0_2-Hooke-taumax15-S0_5-End2-Last1.pdf}
{N100-tau_b0_2-Hooke-taumax15-S0_5-End2-ST2.pdf}
{N100-tau_b0_2-Hooke-taumax15-S0_5-End2-Energy1.pdf}
 \caption{Wavy dispersion error and decrease by dissipative error for
the Hooke system when \m { \qC = 1/2 }, with \m { \qN = 100 }.}
 \label{figHdd}
 \end{figure}

\begin{figure}[ht]
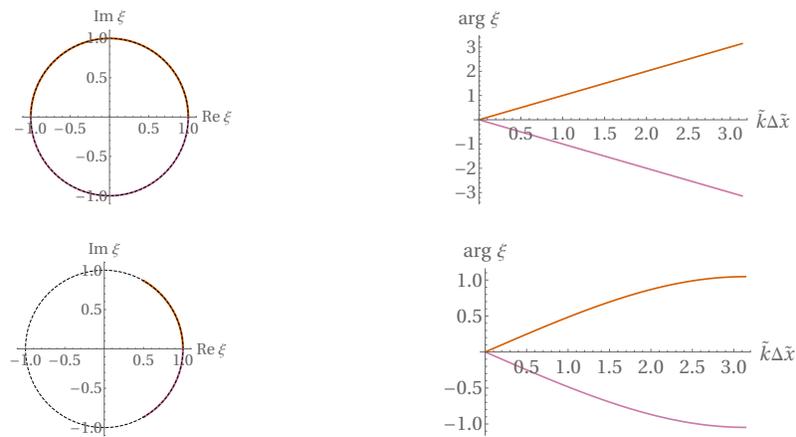
  %% height=20ex
\igg{height=\qFig}{H-ReIm-Courant1.pdf}{H-arg-Courant1.pdf}
 \par\vspace{2.5 ex}
\igg{height=\qFig}{H-ReIm-Courant0_5.pdf}{H-arg-Courant0_5.pdf}
 \caption{\emph{Upper row:} case of \m { \qC = 1 }, \emph{lower row:}
case of \m { \qC = 1/2 }. \emph{Left:} the two roots \m { \qxiiii } in
the complex plane, \emph{right:} \m { \qk } dependence of the argument
of \m { \qxiiii }.}
 \label{figHreimarg}
 \end{figure}

The Hooke system might appear as a simple introductory task for
numerics. This is actually far from true. Already the Hooke case
displays both dissipation error and dispersion error if not treated with
appropriate care \1 1 {see Section~\ref{COMSOL} below, as well as
\cite{JVKR}}. While the greatest danger, instability, is about
exponential exploding of error, dissipation error is `the opposite':
when the signal decreases in time, losing energy due to numerical
artefact only. This type of error is related to \m { \9 6 { \qxi } < 1 }
modes, which indicates that one should try to stay on the unit circle
with \m { \qxi }. On the other side, in addition to the modulus of \m {
\qxi }, its argument can also cause trouble: if \m { \arg \qxi } is not
linear in \m { \qk } then dispersion error is induced, which is
observable as unphysical waves generated numerically around signal
fronts. These errors are present even in a symplectic scheme like ours,
as illustrated in Figure~\ref{figHdd}. More insight is provided by
Figure~\ref{figHreimarg}.

\subsection{\PTZ\ case}

\begin{figure}[ht]
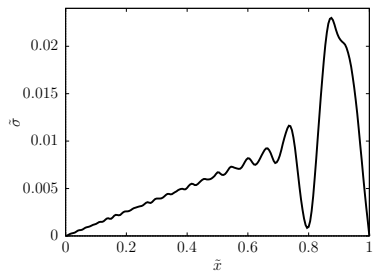
  %% height=20ex
\ig{height=\qFIG}
{N200-tau_b0_2-a0_5-tau_C5-RC0_25-taumax3_6-S0_5-End2-Y0_024.pdf}
 \caption{The stress signal provided by the scheme with \m { \qCC = 1/2
}, \m { \qN = 200 }, for comparison with
% the \m { \qalp = 1/2 } cases in
Figure%
%s~\ref{figPTZ}--
~\ref{figPTZZ}. \1 1 {All other settings are the same as there.}}
 \label{figpd}
 \end{figure}

In case of a dissipative system like the \Red{PTZ} one, it is hard to detect
the dissipative error, \ie to distinguish it from the physical
dissipation. The dispersion error remains visible, as Figure~\ref{figpd}
shows.

Usually, one would need to set \m { \qDt } to be much smaller than \m {
\qtau }, \m { \qtauu } \1 1 {and \m { \qtaub }} to obtain a physically
acceptable approximation. Rewriting the coefficients \re{a0123} as
 \begin{align}  \label{a0123ordo}
\qai{0} & =
\f{ \qalp \f{\qDt}{\qtau} - 1 }{ \0 1 {1 - \qalp} \f{\qDt}{\qtau} + 1 }
= -1 + \Ordo \0 1 { \f{\qDt}{\qtau} } ,
 \qquad
\qai{1} =
3 - \f { 2 \f{\qDt}{\qtau} - 4 \9 1 {\qalp \f{\qDt}{\qtau} -
\f{\qtauu}{\qtau} } \qC^2 \qS^2 }{ \0 1 {1 - \qalp} \f{\qDt}{\qtau} + 1 }
= 3 - 4 \qCC^2 \qS^2 + \Ordo \0 1 { \f{\qDt}{\qtau} },
 \nonumber \\ &
 \\  \nonumber
\qai{2} & =
-3 + \f { \f{\qDt}{\qtau} + 4 \9 2 { \0 1 {1 - \qalp} \f{\qDt}{\qtau} +
\f{\qtauu}{\qtau} } \qC^2 \qS^2 }{ \0 1 {1 - \qalp} \f{\qDt}{\qtau} + 1 }
= -3 + 4 \qCC^2 \qS^2 + \Ordo \0 1 { \f{\qDt}{\qtau} } ,
 \qquad
\qai{3} = 1 ,
 \end{align}
in the limit \m { \f{\qDt}{\qtau} \to 0 }, the characteristic polynomial
reduces to
 \begin{align} \label{polxi2}
\qxi^3 + \9 1 { -3 + 4 \qCC^2 \qS^2 } \qxi^2 + \9 1 { 3 - 4 \qCC^2 \qS^2 }
\qxi - 1 = \9 1 { \qxi - 1 } \9 2 { \qxi^2 + \9 1 { -2 + 4 \qCC^2 \qS^2 }
\qxi + 1 } ,
 \end{align}
with roots satisfying \m { \qxi_0 = \9 6 { \qxii } = \9 6 { \qxiii } = 1
}, excluding thus dissipation error. Especially simple and distinguished
is the case \m { \qCC = 1 }, when the roots are
 \begin{align}  \label{pc1}
\qxi_0 = 1 , \qquad \qxiiii = \e^{ \pm \ii \qk \qDx } ,
 \end{align}
providing dispersion relations linear in \m { \qk } and, hence, getting
rid of dispersion error as well.

\begin{figure}[H]
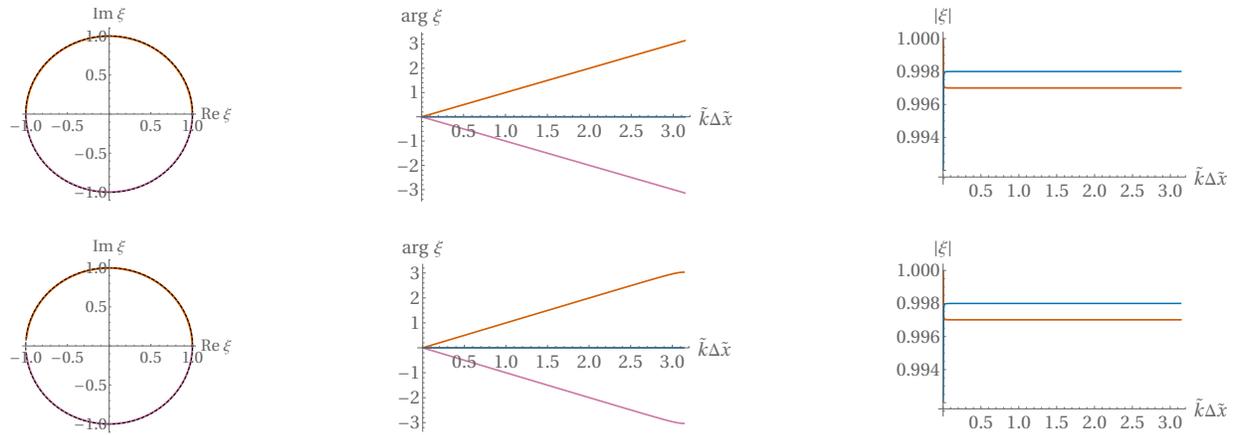
  %% height=20ex
\iggg{height=\qFig}{PT-ReIm-Courant1-a0_5.pdf}
{PT-arg-Courant1-a0_5.pdf}{PT-abs-Courant1-a0_5.pdf}
 \par\vspace{2.5 ex}
\iggg{height=\qFig}{PT-ReIm-Courant1-a0.pdf}
{PT-arg-Courant1-a0.pdf}{PT-abs-Courant1-a0.pdf}
 \caption{Visualization of the three branches \m { \qxi_0 \0 1 {\qk},
\qxii \0 1 {\qk}, \qxiii \0 1 {\qk} } for \m { \qCC = 1 }.
\emph{Upper row:} \m { \qalp = 1/2 }, \emph{lower row:} \m { \qalp = 0
}.}
 \label{figpa}
 \end{figure}

\begin{figure}[H]
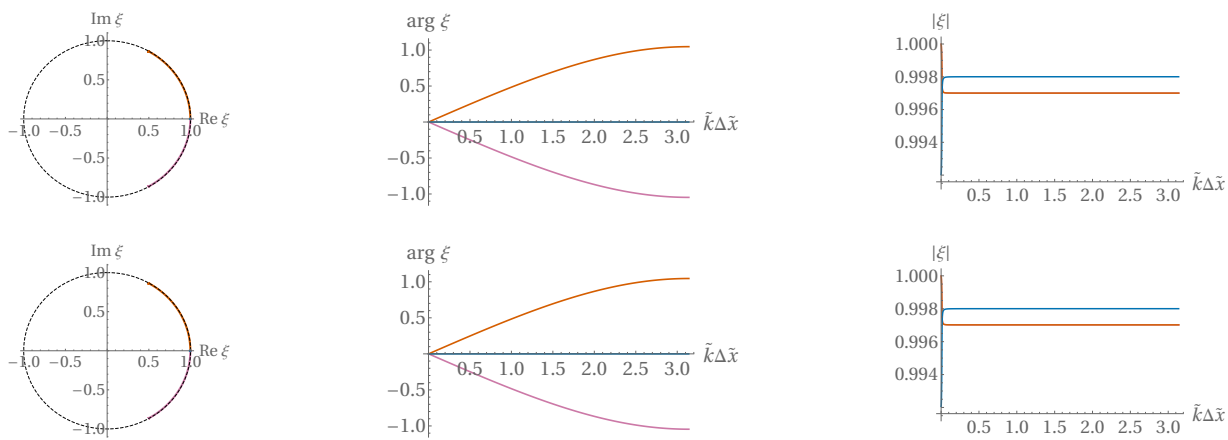
  %% height=20ex
\iggg{height=\qFig}{PT-ReIm-Courant0_5-a0_5.pdf}
{PT-arg-Courant0_5-a0_5.pdf}{PT-abs-Courant0_5-a0_5.pdf}
 \par\vspace{2.5 ex}
\iggg{height=\qFig}{PT-ReIm-Courant0_5-a0.pdf}
{PT-arg-Courant0_5-a0.pdf}{PT-abs-Courant0_5-a0.pdf}
 \caption{Same as Figure~\ref{figpa} but with \m { \qCC = 1/2 }.}
 \label{figpb}
 \end{figure}

\begin{figure}[H]
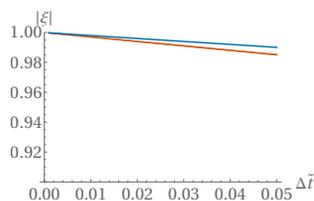
  %% height=20ex
\ig{height=\qFig}{PT-Dtabs-Courant1-a0_5-kDxpiper4.pdf}
 \caption{In Figures~\ref{figpa}--\ref{figpb}, the roots are not exactly
on the unit circle -- here, \m { \qDt } dependence of \m { \9 6 { \qxi_0
} } and \m { \9 6 { \qxiiii } } is displayed, at a neutral value \m {
\qk \qDx = \pi/4 }, for  \m { \qCC = 1 } and \m { \qalp = 1/2 }.}
 \label{figpc}
 \end{figure}

With slightly nonzero \m { \f{\qDt}{\qtau} }, these nice properties get
detuned but only up to \m { \Ordo \0 1 { \f{\qDt}{\qtau} } }, as shown
in Figures~\ref{figpa}--\ref{figpc} \1 1 {prepared at dimensionless time
step value \m { 0.01 }; the detuning appears weaker for \m { \qalp = 1/2
} than for \m { \qalp = 0 }}.

\section{Solutions using the finite element software COMSOL}
  \label{COMSOL}

Finally, for comparative reasons, we present solutions obtained via a
commercial finite element software, namely, COMSOL v5.3a. We considered
the Hookean case, for which the COMSOL implementation is straightforward
since the built-in mathematical environment offers the possibility to
solve such classical partial differential equations, too. 

For the finite element realization, we chose the displacement field as
the primary field variable. Then velocity, plotted in the figures below,
is obtained by taking its time derivative.

To be in conform with the units used for defining the dimensionless
quantities, we set both the propagation speed and the sample length to
unity.  The spatial domain consisted of 100 elements, obtained using the
options of `physics-controlled mesh' and `extremely fine' element size.
On the boundaries, the gradients were prescribed, and the excitation was
given analogously to our above simulations \1 2 {see \re{pulse} and
Figure~\ref{figpulse}}. We examined the schemes for two different pulse
lengths, $\qndtaub = 0.2$ and $\qndtaub = 0.04$.

In what follows, we tested 5 different settings for time stepping, in
order to find the appropriate ones and to compare their effectiveness.
For the simulations, we used a configuration of an i7-7700 CPU with
clocking 3.6 GHz and 16 GB RAM. COMSOL supports parallel computing,
which option has been exploited. Although the run time strongly depends
on other factors, too, it provides a good picture for comparing the
effectiveness of the commercial approach and the scheme presented in
this paper. 

\begin{figure}[H]
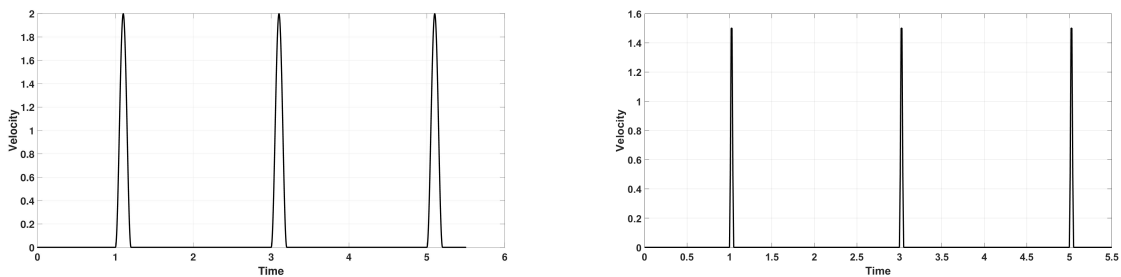

\igg{width=8cm}{hooke_02.jpg}{hooke_004_v2.png}
 \caption{Applying the proposed scheme in case of two different pulse
lengths; $\qndtaub = 0.2$ \1 1 {left} and $\qndtaub = 0.04$ \1 1
{right}. The dimensionless space and time steps are $\Delta \nd{\qx} =
\Delta \nd{\qt} = 0.01$. This time step is actually not much smaller
than the shorter pulse length so, for example, the tips cannot be plotted
accurately when $\qndtaub = 0.04$ but the solution still performs
well.}
 \label{symps1}
\end{figure}

Our scheme,
% \textcolor{red}
{using the same number of spatial elements and time
interval, ran in around 0.2 seconds in Matlab
% in both cases
(using 1 core only) for both pulse lengths, as measured by Matlab.
First, we present the results of our scheme \1 1 {see
Figure~\ref{symps1}}.
 \OMIT{%
 It is easy to observe that
the initial settings in the first case ($\qndtaub = 0.2$) are adequate
and provide a solution without dissipation and dispersion. Although the
second solution is also free from these artificial errors (see the
constant amplitude), the more accurate solution requires a finer
resolution of the pulse. Then, using $\Delta \hattt x = \Delta \hattt t =
0.003$, we can restore the accuracy of the solution quickly, still
keeping the run time around 0.2 seconds (see Figure~\ref{symps2}). }
 }%(V)OMIT
The solutions are apparently free of dissipation and of dispersion.

 \OMIT{%
\begin{figure}[H]
\ig{width=8cm}{hooke_004.jpg}{}
 \caption{Applying the proposed scheme in case of $\qndtaub = 0.04$. The
dimensionless space and time steps are $\Delta \nd{\qx} = \Delta
\nd{\qt} = 0.003$.}
\label{symps2}
\end{figure}
 }%(V)OMIT

Next, we present the outputs obtained via COMSOL used with various
settings.

\subsection{Backward Differentiation Formula (BDF), order 2 and order 5}

\begin{figure}[H]
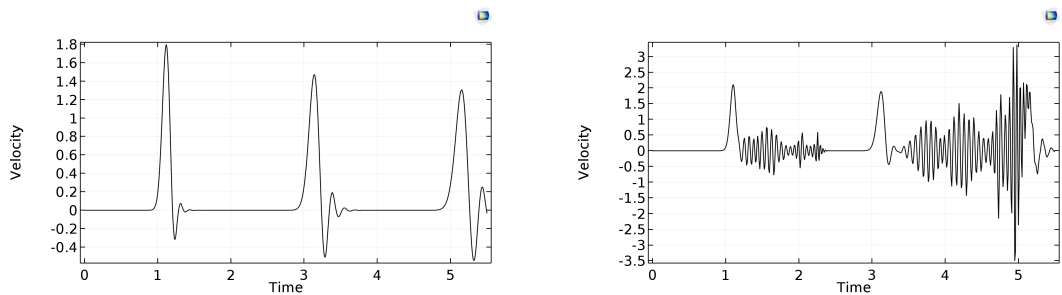
 \igg{width=\qCOMSOL}{BDF_v2.png}{BDF_v1.png}
 \caption{Rear-side velocity history in time for $\qndtaub = 0.2$, with
maximum BDF order being 2 \1 1 {left} and 5 \1 1 {right},
respectively.}
 \label{BDF1}
\end{figure}

In its simplest version, it is the Backward Euler scheme that has good
stability properties, with artificial damping effects. As shown by the
comparison in Figure~\ref{BDF1}, artificial damping is stronger for the
lower order version (\ie maximum BDF order is 2) while, with higher
order schemes, the damping is less significant, therefore, the
artificial oscillations are less suppressed. The run time is between
30--40 seconds.

\subsection{Runge--Kutta-based schemes: Cash--Karp 5}

This scheme results in unstable solutions, independently of the
corresponding settings (initial time step, time step control, and
stiffness control).

\subsection{Runge--Kutta-based schemes: Dormand--Prince 5}

\begin{figure}[H]
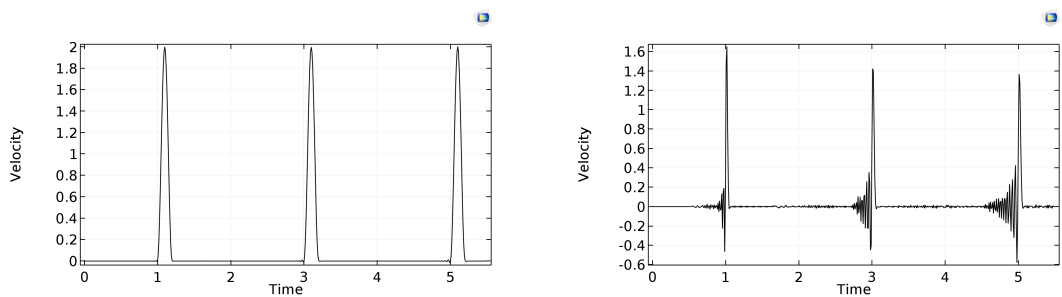

\igg{width=\qCOMSOL}{DP_v1.png}{DP_v2.png}
 \caption{Rear-side velocity history in time, using the Dormand--Prince
time stepping method (left: $\qndtaub=0.2$, right: $\qndtaub=0.04$).}
\label{DP1}
\end{figure}

With this scheme, the numerical stability of the solution strongly
depends on the settings of
% \textcolor{red}
{the} maximum step size growth ratio and the step
size safety factor. At default settings, the solution is unstable. Using
0.1 for the step size safety factor and 0.01 for the maximum step size,
the results can be seen in Figure~\ref{DP1}. Only small oscillations are
observable at the wave front, however, the computation requires almost 2
GB RAM. The run time is strongly influenced by the pulse length. For
$\qndtaub=0.2$, it runs at around 300--320 seconds, using all available
computing capacity. Meanwhile, for the shorter pulse length
$\qndtaub=0.04$, it needs more than 580 seconds. In addition, for
smaller pulse length, dispersive and
%\textcolor{red}
{dissipative} errors are also visible.

\subsection{Runge--Kutta-based schemes: RK34}

\begin{figure}[H]
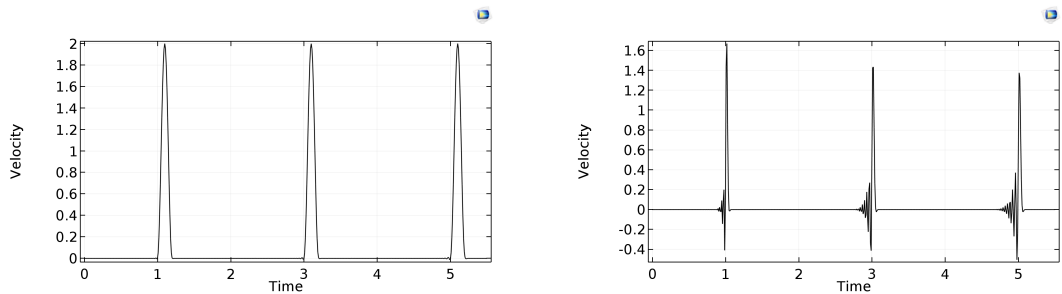

\igg{width=\qCOMSOL}{RK34_v1.png}{RK34_v2.png}
\caption{Rear-side velocity history in time, using the RK34 time
stepping method (left: $\qndtaub=0.2$, right: $\qndtaub=0.04$).}
\label{RK1}
\end{figure}

Using stiffness detection, this scheme solves the problem in the fastest
and most efficient way. However, when the pulse length is
$\qndtaub=0.04$ then both its damping and dispersive properties become
apparent (see Figure~\ref{RK1}).
%With this method, the usual run time was between 40--50 seconds
With this method, the run time was around 45 seconds.

\begin{figure}[H]
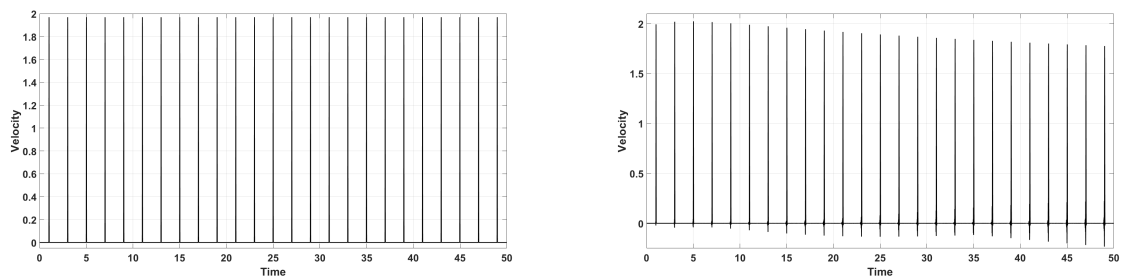

%\igg{width=8cm}{RK34vsSymp1.png}{RK34vsSymp2.png}
 \igg{width=8cm}{Symplectic_longrun.png}{RK34_longrun.png}
\caption{
Rear-side velocity history in time, for pulse length $\qndtaub=0.04$, with
300 nodes. Left: solution by our scheme, right: COMSOL RK34 result.}
 \label{RK1symp}
\end{figure}

Since this COMSOL option proved the best,
in order to test the mesh dependence of its solution, we have examined
the $\qndtaub=0.04$ case with 300 space cells \1 1 {\m{\Delta \nd{\qx} =
\Delta \nd{\qt} = 0.0033}} as well, for a longer process \1 1 {100
bounces}. With these settings, our scheme required 0.3 seconds in Matlab
and shows no numerical artefact while the COMSOL solution took 9649
seconds and exhibits apparent dissipative error and mild but increasing
dispersion error around the rear of the pulse \1 1 {see
Figure~\ref{RK1symp}}.
 \OMIT{%
% \textcolor{red}
{using 100 nodes. Moreover, this method is also tested
for finer mesh when $\qndtaub=0.04$, using 300 nodes, in order to
compare directly this method to the proposed finite difference scheme
since the RK34 seems to be the best available option in COMSOL. In this
modified case, the run time was between 130--150 seconds.}
 }%(V)OMIT

To summarize, compared to our scheme realized in Matlab, COMSOL run
times are 100--1000--10000 times larger, with large memory demand, and various
settings have to be tuned to obtain a stable solution with moderate
artificial dissipation and dispersion.

%qqqqwwww

\section{Discussion}

Choosing a good finite difference numerical scheme for a continuum
thermodynamical problem is not easy. A good starting point can be a
symplectic scheme for the reversible part, as done here, too. Another
advantage is provided by a \Red{staggered} arrangement of quantities by half
space and time steps, suited to balances, to the kinematic equations, to
the Onsagerian equations \etc

Even with all such preparations, instability is a key property to
ensure. And when all these are settled, dissipative and dispersive
errors can invalidate our calculation, which may not be recognized when
the continuum system is dissipative and when it allows wave
propagation as well.

Notably, there is a principal difference between the stability problems
of a numerical method and the stability issues for a continuum
phenomenon itself. The former
% emerge due to
 are induced by
the approximations and depend on the type of approximation, meanwhile
the original continuum system may be fine regarding stability -- for
example, ensured by thermodynamical
% \Red{consistency}.  %% !!! overruling Grammarly: Mac
%% Dictionary: Am. English: consistence is another term for consistency
 consistence.
It is interesting to
realize that, nevertheless, these two types of instability are not
completely independent. In one of the directions, the stability
investigation of a numerical scheme may provide information for the
underlying continuum phenomenon as well. An example for this has been
provided by our \re{tt} above, which is a condition that is independent
of the time step, the space step, the parameter \m { \qalp } that
parametrizes the scheme, and any other aspect of the scheme. Rather, it
is a condition on the underlying continuum model. In the present case,
we already know this condition as one of the stability requirements
imposed by thermodynamical consistency, seen at \re{tauhat}. This
example enlights that, in more complicated problems, it is also worth
investigating the stability conditions of the numerical method and
trying to distill \emph{scheme independent information} on the
continuum system from them.

In parallel, the other direction is when
% \Red{the stability}  %% !!! overruling Grammary
 stability
of a continuum system
can be used \emph{to devise stable numerical methods}. One such example
is the case of symplectic numerical schemes, which are actually exact
integrators of a certain Hamiltonian system -- a slightly different one
from the original system. The generalization of this way of thinking to
nonconservative systems is a promising research direction.

Concerning the future prospects of the study provided here, the findings
can be supplemented by comparison with analytical solutions and further
finite element calculations, performed for the whole \Red{PTZ} system.

Another logical continuation of the present line of research is
extension of the scheme to 2D and 3D space -- this is actually work in
progress \cite{JVKR}.

Regarding the thermodynamical system to be investigated, the whole
Kluitenberg--Verh\'as family -- which the present \Red{PTZ} model is a
subclass of -- could be studied. The presence of second derivative of
strain, and actually already the Kelvin--Voigt subfamily, brings in the
aspect of parabolic characteristics so useful implications may be gained
for other thermodynamical areas like non-Fourier heat conduction as
well.

%In conclusion, reliable
Reliable numerical methods for thermodynamical systems,
which avoid all the various pitfalls, are an important direction for
future research.

\vspace{6pt} 

 \OMIT{%
\authorcontributions{For research articles with several authors, a short
paragraph specifying their individual contributions must be provided.
The following statements should be used ``conceptualization, X.X. and
Y.Y.; methodology, X.X.; software, X.X.; validation, X.X., Y.Y. and
Z.Z.; formal analysis, X.X.; investigation, X.X.; resources, X.X.; data
curation, X.X.; writing--original draft preparation, X.X.;
writing--review and editing, X.X.; visualization, X.X.; supervision,
X.X.; project administration, X.X.; funding acquisition, Y.Y.'', please
turn to the 
\href{http://img.mdpi.org/data/contributor-role-instruction.pdf}{CRediT
taxonomy} for the term explanation. Authorship must be limited to those
who have contributed substantially to the work reported.}
 }%(V)OMIT

 \funding{%The research reported in this paper was supported
The work was supported by the grants National Research, Development and
Innovation Office -- NKFIH 116375 \& 116197, NKFIH KH 130378, and NKFIH
K124366(124508), by FIEK-16-1-2016-0007, and by the National Research,
Development and Innovation Fund (TUDFO/51757/2019-ITM), Thematic
Excellence Program.}

\acknowledgments{The authors thank Hans Christian \"Ottinger, Miroslav
Grmela, Mikl\'os Mincsovics, and Merc\'edesz Vass for stimulating
discussions and suggestions for literature.}

\conflictsofinterest{The authors declare no conflict of interest.}

%%%%%%%%%%%%%%%%%%%%%%%%%%%%%%%%%%%%%%%%%%
%% optional
\appendixtitles{no} %Leave argument "no" if all appendix headings stay EMPTY (then no dot is printed after "Appendix A"). If the appendix sections contain a heading then change the argument to "yes".
%\appendix
%\section{}
%\unskip
%\subsection{}
% The appendix is an optional section that can contain details and data
% supplemental to the main text. For example, explanations of
% experimental details that would disrupt the flow of the main text, but
% nonetheless remain crucial to understanding and reproducing the
% research shown; figures of replicates for experiments of which
% representative data is shown in the main text can be added here if
% brief, or as Supplementary data. Mathematical proofs of results not
% central to the paper can be added as an appendix.

%\section{}
%All appendix sections must be cited in the main text. In the appendixes, Figures, Tables, etc. should be labeled starting with `A', e.g., Figure A1, Figure A2, etc. 

\reftitle{References}
% Please provide either the correct journal abbreviation
% http://www.issn.org/services/online-services/access-to-the-ltwa/) or
% the full name of the journal.

\end{document}